\documentclass[a4paper,10pt]{article}
\usepackage[a4paper, margin=2cm]{geometry}

\usepackage[utf8]{inputenc}
\usepackage[T1]{fontenc}

\usepackage{lmodern}

\usepackage{amsmath}
\allowdisplaybreaks
\usepackage{amsfonts}
\usepackage{amssymb}
\usepackage{dsserif}
\usepackage{bm}
\usepackage{mathrsfs}
\usepackage{booktabs}
\usepackage{graphicx}
\usepackage{url}
\usepackage{xcolor}

\usepackage[normalem]{ulem}
\makeatletter
\newcommand*{\transpose}{%
	{\mathpalette\@transpose{}}%
}
\newcommand*{\@transpose}[2]{%
	\raisebox{\depth}{$\m@th#1\intercal$}%
}
\makeatother

\usepackage{accents}

\usepackage{amsthm}

\makeatletter
\def\th@remark{%
	\thm@headfont{\bfseries}%
	\normalfont 
	\thm@preskip\topsep \divide\thm@preskip\tw@
	\thm@postskip\thm@preskip
}
\makeatother

\theoremstyle{remark}
\newtheorem*{remark}{Remark}

\usepackage[english]{babel}
\usepackage{csquotes}

\usepackage[hidelinks]{hyperref}

\usepackage[numbers]{natbib}
\usepackage{doi}

\title{Nonlinear dispersive waves in soft elastic laminates\\ under finite magneto-deformations}
\author{Harold Berjamin \textsuperscript{a}, Stephan Rudykh \textsuperscript{a}\\
	{\footnotesize
		\begin{tabular}{l}
			~ \\
			\textsuperscript{a}School of Mathematical and Statistical Sciences, University of Galway, University Road, Galway, Republic of Ireland
	\end{tabular}}
}
\date{}

\begin{document}
	
\maketitle
	
\begin{abstract}
	\noindent
	Layered media can be used as acoustic filters, allowing only waves of certain frequencies to propagate. In soft magneto-active laminates, the shear wave band gaps (i.e., the frequency intervals for which shear waves cannot propagate) can be adjusted after fabrication by exploiting the magneto-elastic coupling. In the present study, the control of shear wave propagation in magneto-active stratified media is revisited by means of homogenisation theory, and extended to nonlinear waves of moderate amplitude. Building upon earlier works, the layers are modelled by means of a revised hard-magnetic material theory for which the total Cauchy stress is symmetric, and the incompressible elastic response is of generalised neo-Hookean type (encompassing Yeoh, Fung-Demiray, and Gent materials). Using asymptotic homogenisation, a nonlinear dispersive wave equation with cubic nonlinearity is derived, under certain simplifying assumptions. In passing, an effective strain energy function describing such laminates is obtained. The combined effects of nonlinearity and wave dispersion contribute to the formation of solitary waves, which are analysed using the homogenised wave equation and a modified Korteweg--de Vries (mKdV) approximation of the latter. The mKdV equation is compared to direct numerical simulations of the impact problem, and various consequences of these results are explored. In particular, we show that an upper bound for the speed of solitary waves can be adjusted by varying the applied magnetic field, or by modifying the properties of the microstructure.
	
	\medskip
	
\end{abstract}


\medskip

\section{Introduction}

Multi-layered or stratified materials have found many applications in Engineering, starting with sandwich panels used in construction and manufacturing \citep{allen69}. Following the early developments of 3D printing in the 1980s, soft laminates have also been employed as a building block for additive manufacturing processes such as deposition techniques \citep{polymers22}. Stratified media made of soft layers are also found in nature, for example in the form of animal arteries, skin, and other biological tissues. Beyond skin wrinkling and aging \citep{alawiye19,zhao20}, the study of the skin's structure and of its mechanical properties is useful for the prediction of injuries, as well as for the development of surrogates and artificial skins \citep{mahoney18,chen19}.

Magneto-rheological elastomers are soft rubber-like materials filled with magnetic particles. In contrast with purely elastic materials, their mechanical response is sensitive to external magnetic fields. Essentially, they can be separated into two main groups depending on the type of magnetic particles that were inserted in the rubber-like solid. Here, we restrict our attention to so-called \emph{hard}-magnetic particles, which exhibit a residual magnetisation in the absence of an external magnetic field \citep{zhao19,lucarini22}, in opposition to \emph{soft}-magnetic particles. By design, such active materials are excellent candidates for the development of controllable structures, which are of great interest for potential applications in robotics and material science. In this context, the properties of magneto-active materials can also be exploited for the development of bio-inspired sensors \citep{liu24}.

Periodic (bi)-laminates consisting of two layers of alternating materials are the simplest type of composite material. They are also the simplest type of phononic crystal or of acoustic metamaterial, which exhibits a periodic variation of its mechanical properties along one dimension only, known as the lamination direction. One of their main characteristics is the possibility of having a phononic \emph{band gap}. In this case, waves of selected frequency ranges are prevented from being transmitted through the material, due to destructive interferences \citep{deymier13}. Thus, these simple materials with a periodic micro-structure can be used for wave and vibration filtering. Under some conditions, the filtering properties of the material can be adjusted after fabrication, for instance based on the application of a controlled mechanical force \citep{parnell07}.

The study of magneto-active bi-laminates has been approached by several authors. \citet{rudykh13} studied the stability of magneto-active laminates filled with soft-magnetic particles, whereas \citet{karami19} investigated linear shear wave propagation transversely to the layers. One of the features demonstrated therein is the tunability of the band gaps, i.e., the fact that the wave filtering properties of the material can be adjusted by varying the magnitude of an external magnetic field. These works were later extended to hard-magnetic laminates \citep{zhang22,alam23}, as well as to shear waves with an arbitrary direction of propagation \citep{alam25}. Further instability results are presented by \citet{yao24}, while viscoelastic dissipation is accounted for in recent works \citep{wang24,ruggieri25}.

To the authors' present knowledge, the description of the nonlinear time-dependent motion of soft magneto-active layered materials remains to be explored. To bridge this gap, the present study is dedicated to the propagation of nonlinear shear waves of moderate amplitude in such periodic media. The main step of this analytical and computational study is the derivation of an effective wave equation which accounts for nonlinearity and wave dispersion. Here, we rely on asymptotic homogenisation to derive the effective wave equation based on the properties of the individual layers \citep{andrianov13}. Then, several properties of the nonlinear dispersive waves are described by means of analytical approximations as well as numerical simulations.

The manuscript is organised as follows. In Section~\ref{sec:Constitutive}, we recall the governing equations of nonlinear magneto-elasticity, and we describe our constitutive model. The latter is a revised version of hard-magnetic material theories for which the total stress tensor satisfies the usual symmetry requirements \citep{dorfmann24}. Then, under the incompressibility assumption, we use this theoretical framework to study the deformation of one-dimensional composite media made of such magneto-active layers, as well as the wave propagation characteristics in these layers (Section~\ref{sec:Laminate}). For a large class of soft solids (including pre-deformed magneto-active ones), shear waves propagating along the lamination direction are governed by the nonlinear wave equation
\begin{equation}
	\left(g^{(\alpha)} + h^{(\alpha)} u_y^2\right) u_{yy} = \rho^{(\alpha)} u_{tt} ,
	\label{WaveAlpha}
\end{equation}
where $u$ is the shearing displacement, and the subscript notation $u_y$ and $u_t$ indicates partial differentiation with respect to space and time, respectively. Here, $\rho^{(\alpha)}$ is the mass density, and the coefficients $g^{(\alpha)}$, $h^{(\alpha)}$ are constant stiffness parameters. The latter can be potentially adjusted based on an applied (magneto)-deformation.

\begin{table*}
	\caption{Effective physical parameters governing shearing motions \eqref{WaveEff} in a bi-laminate. The motion is governed by Eq.~\eqref{WaveAlpha} in each phase labelled by $\alpha = 1,2$, and $\nu^{(\alpha)}$ denotes the volume fraction of the constituent $\alpha$. The effective wave speed satisfies $\mathfrak{c}^2 = \langle g \rangle / \langle \rho \rangle $. \label{tab:Effective}}
	\vspace{0.5em}
	\centering
	{\renewcommand{\arraystretch}{1}
	\small
	\begin{tabular}{cccc}
    \toprule 
    $\langle g \rangle$ & $\langle \rho \rangle$ & $\zeta$ & $\eta$ \\
    \midrule
    $\displaystyle \left( \frac{\nu^{(1)}}{g^{(1)}} + \frac{\nu^{(2)}}{g^{(2)}} \right)^{-1}$ & $\nu^{(1)} \rho^{(1)} + \nu^{(2)} \rho^{(2)}$ & $\displaystyle \langle g \rangle^{3} \left(\frac{\nu^{(1)} h^{(1)}}{(g^{(1)})^4} + \frac{\nu^{(2)} h^{(2)}}{(g^{(2)})^4}\right)$ & $\displaystyle \frac{\mathfrak{c}^{4}}{12} \frac{(\nu^{(1)} \nu^{(2)})^2}{(g^{(1)} g^{(2)})^2} \left( \rho^{(1)}g^{(1)} -\rho^{(2)}g^{(2)} \right)^2$ \\ 
    \bottomrule 
    \end{tabular}}
\end{table*}

In Section~\ref{sec:Homogen}, we use these results to derive an effective theory for wave propagation in nonlinear magneto-active laminates, using the two-scale homogenisation technique \citep{andrianov13}. In a perfectly bonded and periodic bi-laminate made of such materials ($\alpha = 1, 2$ in Eq.~\eqref{WaveAlpha}), we show that the shearing motion is governed by the partial differential equation
\begin{equation}
	\mathfrak{c}^2 \left(1 + \zeta u_y^2 \right) u_{yy} +  \eta \ell^2 \mathfrak{c}^2 u_{yyyy} = u_{tt} ,
	\label{WaveEff}
\end{equation}
where $\mathfrak{c}$ is the effective shear wave speed in the laminate. The effective shear modulus $\langle g \rangle = \langle \rho \rangle \mathfrak{c}^2$ and the effective mass density $\langle \rho \rangle$ describe the laminate's linear and non-dispersive response, see Table~\ref{tab:Effective} for detailed expressions of the effective parameters. Here, a physical quantity in brackets $\langle \cdot \rangle$ represents the effective macroscopic counterpart of that quantity. Hence, we could introduce $\langle h \rangle = \zeta \langle g \rangle$, in a similar fashion. The dimensionless parameters $\zeta$ and $\eta$ account for nonlinearity and wave dispersion, respectively, and the parameter $\ell$ is the microstructure's characteristic length (spatial period), see Figure~\ref{fig:Struct}b. Using homogenisation theory, we then connect the effective wave equation \eqref{WaveEff} to nonlinear elastic models of soft solid. More specifically, we derive an anisotropic strain energy function \eqref{Weff} for soft layered materials which is consistent with Eq.~\eqref{WaveEff}. Here, we assume that the layers are described by the two-term Yeoh model, which is viewed as a small perturbation of the neo-Hookean theory of elasticity.

In Section~\ref{sec:NLWaves}, formulations of \eqref{WaveEff} with alternative dispersion terms and consequences of this result are investigated.  Here, we cover linear wave dispersion properties, unidirectional nonlinear waves described by a modified Korteweg--de Vries (mKdV) equation, as well as soliton-like nonlinear wave solutions. Furthermore, we validate the mKdV model against direct numerical simulations. In Section~\ref{sec:Tune}, we present various consequences of the tunability of the coefficients $g^{(\alpha)}$ and $h^{(\alpha)}$ based on an applied magneto-deformation. In particular, we show that the dispersion properties of small-amplitude waves can be adjusted by exploiting the magneto-elastic coupling or by modifying the characteristics of the microstructure, and a similar observation is made with regard to the maximum velocity of solitary wave solutions.

\section{Governing equations}\label{sec:Constitutive}

\subsection{Theoretical framework}

We follow \citet{dorfmann04} and introduce the deformation gradient tensor $\bm{F} = \partial{\bm x}/\partial{\bm X}$, which is the gradient of the map that connects the position ${\bm x} = {\bm \chi} ({\bm X}, t)$ of a material point in the current configuration to its position $\bm X$ in the reference configuration. The volume dilation corresponds to the determinant of the deformation gradient tensor,
\begin{equation}
	J = \det {\bm F} = \rho_0/\rho > 0 ,
	\label{VolumeChange}
\end{equation}
where $\rho$ denotes the mass density, and $\rho_0$ is its initial counterpart defined in the undeformed configuration. We use overdots to denote the material time derivative $\partial/\partial t$ at constant $\bm X$, which satisfies
\begin{equation}
	\dot{J}/J = \text{tr}\, {\bm L} = -\dot \rho/\rho . 
	\label{VolumeRate}
\end{equation}
Here, $\bm{L}$ is the Eulerian velocity gradient tensor, which can be expressed by $\dot{\bm F}\bm{F}^{-1}$ in the material description of the motion. The above identity expresses the balance of mass.

For an isolated magneto-elastic solid, the mechanical equations of motion in the spatial description read
\begin{equation}
    \text{div}\, \bm{\sigma} = \rho \bm{a} , \quad \bm{\sigma} = \bm{\sigma}^\text{T} ,
    \label{Equil}
\end{equation}
where $\text{div}$ is the divergence operator in the current configuration, the exponent \textsuperscript{T} signifies the transpose, $\bm a$ is the acceleration field, and external body forces have been neglected. These equations express the balance of linear momentum and of angular momentum, respectively. The tensor $\bm \sigma$ represents the total magneto-elastic stress tensor (i.e., it encompasses the `mechanical' part of the Cauchy stress tensor and the magnetic body force), whose expression is provided subsequently.

In the absence of a prescribed current density, the Maxwell field equations read
\begin{equation}
    \begin{aligned}
    &\text{curl}\, {\mathbb h} = {\bf 0}, &\quad {\mathbb h} &= \bm{F}^{-\text{T}} {\mathbb H} ,
    \\
    &\text{div}\, {\mathbb b} = 0, &\quad {\mathbb b} &= J^{-1} \bm{F} {\mathbb B} ,
    \end{aligned}
    \label{Maxwell}
\end{equation}
which are Ampère's and Gauss' laws of magnetostatics, respectively. The vectors ${\mathbb b}$ and ${\mathbb h}$ are the magnetic induction field (expressed in T) and the magnetic strength field (in A/m) in the current configuration, whereas $\mathbb B$, $\mathbb H$ are their counterparts in the reference configuration. In addition, we introduce also the magnetisation vector $\mathbb m$ and its Lagrangian counterpart $\mathbb M$, such that
\begin{equation}
	{\mathbb m} = {\mathbb b}/\mu_0 - {\mathbb h} = \bm{F}^{-\text{T}} {\mathbb M} ,
	\label{Magnetisation}
\end{equation}
where $\mu_0 = 4\pi \times 10^{-7}$ N/A\textsuperscript{2} is the magnetic permeability of vacuum.

At an interface with unit normal $\bm n$ between two such materials, the normal tractions $\bm{\sigma} \cdot \bm{n}$ must be continuous, as well as the quantities ${\mathbb h} \times {\bm n}$ and ${\mathbb b}\cdot \bm{n}$ when no surface current density is prescribed at the boundary. In vacuum, the total stress arising from the magnetic field is the Maxwell stress
\begin{equation}
    \bm{\sigma}^\text{m} = \mu_0^{-1} \big({\mathbb b}\otimes {\mathbb b} - \tfrac12 |{\mathbb b}|^2 \bm{I}\big) ,
    \label{StressMax}
\end{equation}
to be used for the boundary conditions at a free interface, where $|{\mathbb b}|^2 = {\mathbb b}\cdot{\mathbb b}$ is the squared norm of $\mathbb b$. The notation $\bm I$ stands for the second-order identity tensor, and $\otimes$ is the tensor product (dyadic, or outer product).

To complete the theory, we need to provide the constitutive relationship that links the stress tensor $\bm\sigma$ to the magnetic quantities $\mathbb b$, $\mathbb h$ and to the kinematic quantity $\bm F$. Such a relationship needs to comply with thermodynamic restrictions. For this purpose we follow the steps described in \citet{kovetz00}, which are coherent with the conventions of the monograph by \citet{holzapfel00} for the componentwise calculation of tensor derivatives and differential operators. The interested reader will find other related derivations in \citet[Chap. 1]{CISM11}.

For a material in thermal equilibrium where the heat flux and heat supply are negligible, the local form of the balance of internal energy and of the entropy inequality read \citep[pp. 219-220]{kovetz00}
\begin{equation}
	\rho \dot{e} = (\bm{\sigma} - \mathbb{h}\otimes \mathbb{b} + (\mathbb{h}\cdot \mathbb{b})\bm{I}) : \bm{L} + \mathbb{h}\cdot \dot{\mathbb{b}} , \quad \rho \dot\eta \geq 0 ,
	\label{Thermo}
\end{equation}
where $e$, $\eta$ are the specific internal energy and the specific entropy. The colon $:$ indicates a double contraction operation between second-order tensors.

Next, we assume that the free energy per unit of reference volume is given by an expression of the form
\begin{equation}
	\rho_0(e - \theta\eta) = \Omega(\theta, \bm{F}, \mathbb{B}) ,
	\label{FreeEner}
\end{equation}
where $\theta > 0$ is the absolute temperature, and $\Omega$ is a potential energy. Thus, upon differentiation of the above with respect to time and multiplication of \eqref{Thermo}\textsubscript{2} by $J\theta$, we arrive at the dissipation inequality
\begin{equation}
	\rho_0\theta \dot\eta = \rho_0 (\dot e - \dot\theta\eta) - \dot \Omega \geq 0 ,
	\label{DissIneq}
\end{equation}
which needs to be satisfied for every magneto-deformation. By the chain rule, the time derivative $\dot\Omega$ can be replaced by the sum $\Omega_\theta \dot\theta + \Omega_{\bm F} : \dot{\bm F} + \Omega_{\mathbb{B}} \cdot\dot{\mathbb{B}}$, where the subscripts are a shorthand abbreviation for partial differentiation.

We describe the dependence of the potential $\Omega$ with respect to $\theta$ implicitly through the Gibbs--Helmholtz equation $\rho_0\eta = -\Omega_\theta$. Upon using \eqref{Maxwell} as well as differentiation rules, we rewrite \eqref{Thermo}\textsubscript{1} in the form $\rho \dot{e} = {\bm \sigma}: {\bm L} + J^{-1}\mathbb{H}\cdot \dot{\mathbb B}$. Hence, multiplying this identity by $J = \rho_0/\rho$ and using \eqref{DissIneq}, the following Clausius--Duhem inequality is obtained \citep{saxena13},
\begin{equation}
	0 \leq \rho_0 \theta \dot{\eta} 
		= (\bm{P} - \Omega_{\bm F}):\dot{\bm F} + (\mathbb{H} - \Omega_{\mathbb B})\cdot \dot{\mathbb B} ,
	\label{Dissipation}
\end{equation}
where $\bm{P} = J\bm{\sigma}\bm{F}^{-\text{T}} $ is the first Piola--Kirchhoff stress tensor deduced from the total magneto-elastic stress $\bm \sigma$. Based on the above relationships, we note that the dissipation $\rho_0\theta \dot\eta$ equals zero if
\begin{equation}
    \bm{P} = \frac{\partial \Omega}{\partial \bm F} , \quad
    {\mathbb H} = \frac{\partial \Omega}{\partial {\mathbb B}} .
    \label{ConstitutiveDer}
\end{equation}
In isotropic solids, it can be shown that $\Omega$ depends on $\bm F$ only through a dependency with respect to the right Cauchy--Green strain tensor $\bm{C} = \bm{F}^\text{T}\bm{F}$. A specification of the dependence of $\Omega$ with respect to $\theta$ can be avoided if we are in a situation where the temperature variations are negligible (cf. \citet{chadwick74} and related works for the modelling of thermo-elastic effects).

The constitutive relationships \eqref{ConstitutiveDer} can be expressed in terms of different variables and potentials, see for instance \citet[Chap. 3]{CISM11}. If we introduce the function $\phi$ such that
\begin{equation}
	\Omega = \rho_0 \phi (\bm{F}, \mathbb{b}) + \tfrac12 J |\mathbb{b}|^2 / \mu_0 ,
	\label{Steigmann}
\end{equation}
with $\mathbb b$ defined in Eq.~\eqref{Maxwell}, then we recover the expressions provided by \citet{steigmann04} based on the potential energy $\phi$. Similarly, introducing the functions $\Omega^*$, $\psi$ such that
\begin{equation}
	\Omega = \Omega^*(\bm{F}, \mathbb{H}) + \mathbb{H}\cdot \mathbb{B} , \quad \Omega^* = \rho_0 \psi (\bm{F}, \mathbb{h}) - \tfrac12 \mu_0 J |\mathbb{h}|^2 ,
	\label{Legendre}
\end{equation}
we recover the expressions provided by \citet{giorgi25} based on the potential energy $\psi$ (see also \citet{morro24}). The $\bm{F}$-$\mathbb{H}$ formulation used by \citet{mukherjee22} is recovered by using the potential energy $\Omega^*$ introduced above, which is a partial Legendre transformation of $\Omega$.

\begin{remark}
An alternative Clausius--Duhem inequality can be obtained by expressing the internal energy as a function of the form $\rho_0 e = \acute\Omega(\eta, \bm{F}, \mathbb{B})$. Doing so amounts to performing a partial Legendre transformation with respect to the thermal variable, i.e., we have $\Omega = \acute\Omega - \rho_0\theta\eta$. Inserting this expression into Eq.~\eqref{DissIneq} yields
\begin{equation}
	\rho_0\theta \dot\eta = \rho_0 (\dot e + \theta\dot \eta) - \dot {\acute\Omega} \geq 0  .
\end{equation}
Formally, the same constitutive relationships as \eqref{Dissipation}-\eqref{ConstitutiveDer} can then be derived (with $\Omega$ replaced by $\acute\Omega$), based on the Gibbs--Helmholtz relationship $\rho_0\theta = \acute\Omega_\eta$. The explicit dependence of $\acute\Omega$ with respect to $\eta$ can be dropped as long as we are in a situation where the entropy variations are negligible. Finally, the same transformations \eqref{Steigmann}-\eqref{Legendre} can be performed as for $\Omega$.
\end{remark}

\subsection{Hard-magnetic soft solids}

Hard-magnetic soft solids have the ability to retain a high magnetisation in the absence of an external magnetic field, the remnant magnetisation. In ideal hard-magnetic soft solids, the magnetic permeability $\mu$ is assumed to be close to that of vacuum (or air), $\mu_0$. In addition, it is assumed that the magnetic induction depends linearly on the magnetic field within a given range, for magnetic fields that are much less intense than the magnetic coercivity \citep{zhao19}.

Based on the above observations, one might decompose the magnetic field $\mathbb h$ according to a linear law \citep{zhang22},
\begin{equation}
	\mathbb{b} = \mathbb{b}^\text{r} + \mu \mathbb{h} = \mu_0 \mathbb{m}^\text{r} + \mu \mathbb{h} ,
	\label{MagLinear}
\end{equation}
where $\mu$ is the magnetic permeability, and the remnant magnetisation $\mathbb{m}^\text{r} = \mathbb{b}^\text{r} / \mu_0$ is acquired after the material has been exposed to a strong magnetising field. Thus, the remnant magnetic induction is such that $\mathbb{b}^\text{r} = \mathbb{b}|_{\mathbb{h}={\bf 0}}$. Furthermore, Eq.~\eqref{Magnetisation} yields the relationships
\begin{equation}
	\mathbb{m} = \mathbb{m}^\text{r} + \frac{\mu \chi}{\mu_0} \mathbb{h} = \frac{1}{\mu} \mathbb{b}^\text{r} + \frac{\chi}{\mu_0}\mathbb{b} ,
	\label{MagRelations}
\end{equation}
where $\chi = (\mu-\mu_0)/\mu$ is the magnetic susceptibility. These expressions recover the equality $\mathbb{m} = \mathbb{m}^\text{r}$ whenever the magnetic permeability equals that of vacuum (i.e., for $\mu = \mu_0$), which corresponds to the law proposed by \citet{zhao19}. The Lagrangian counterparts $\mathbb{B}^\text{r}$ and $\mathbb{M}^\text{r}$ of $\mathbb{b}^\text{r}$, $\mathbb{m}^\text{r}$ are defined in a similar fashion to the vectors $\mathbb{B}$, $\mathbb{M}$ introduced in Eqs.~\eqref{Maxwell}-\eqref{Magnetisation}.

Based on \eqref{MagLinear}, we have
\begin{equation}
	\mathbb{H} = (\mu J)^{-1} \bm{C} (\mathbb{B} - \mathbb{B}^\text{r}),
\end{equation}
where $\bm{C} = \bm{F}^\text{T}\bm{F}$, and $\mathbb{B}^\text{r} = J\bm{F}^{-1} \mathbb{b}^\text{r}$ is the Lagrangian counterpart of $\mathbb{b}^\text{r}$. This expression can be integrated with respect to $\mathbb{B}$ according to Eq.~\eqref{ConstitutiveDer}\textsubscript{2} to arrive at the following expression of the magnetoelastic energy:
\begin{equation}
	\begin{aligned}
	\Omega &= \Omega^\text{e} - \frac{1}{\mu J}\bm{F}\mathbb{B}^\text{r} \cdot \bm{F}\mathbb{B} + \frac1{2\mu J} |\bm{F}\mathbb{B}|^2 , \\
	&= \Omega^\text{e} - \frac{\mu_0}{\mu}\mathbb{M}^\text{r} \cdot \mathbb{B} + \frac1{2\mu J} |\bm{F}\mathbb{B}|^2 , 
	\end{aligned}
	\label{HMAE}
\end{equation}
where $\Omega^\text{e}$ is an energy function independent of $\mathbb B$, and where we have introduced the Lagrangian counterpart of the remnant magnetisation, $\mathbb{M}^\text{r} = (\mu_0 J)^{-1} \bm{C} \mathbb{B}^\text{r}$. For $\mu=\mu_0$, the expression \eqref{HMAE}\textsubscript{2} of the total energy is identical to that in Eqs.~(16)-(17) of \citet{dorfmann24}. As noted therein, the last term in the expression of $\Omega$ was missing in \citet{zhao19}. Within the present formalism, the contribution of the self-magnetisation produced by the magnetised body in the entire space does not appear explicitly (i.e., the theory is purely local). To complete the picture, the interested reader is referred to relevant literature \cite{steigmann04,rahmati23}.

Now, let us use the expression \eqref{HMAE}\textsubscript{1} of $\Omega$ to derive the expression of the stress tensor $\bm\sigma$, in which $\mathbb{B}^\text{r}$ can be viewed as a constant vector. Using Eq.~\eqref{Steigmann} together with the connection
\begin{equation}
	\rho_0 \phi = \Omega^\text{e} - \frac1{\mu} \bm{F}\mathbb{B}^\text{r} \cdot \mathbb{b} - \frac{\chi}{2\mu_0} J |\mathbb{b}|^2 ,\
	\label{HMAEPart}
\end{equation}
we find
\begin{equation}
	\begin{aligned}
		\bm{\sigma} &= J^{-1} \frac{\partial \Omega^\text{e}}{\partial \bm F} \bm{F}^\text{T}
	- \mu^{-1} \mathbb{b} \otimes \mathbb{b}^\text{r} - \frac{\chi}{2\mu_0} |\mathbb{b}|^2 \bm{I} \\
	&\quad - \mathbb{m}\otimes\mathbb{b} + (\mathbb{m}\cdot\mathbb{b})\bm{I} + \bm{\sigma}^\text{m} ,
	\end{aligned}
	\label{StressPart}
\end{equation}
where $\bm{\sigma}^\text{m}$ is defined in Eq.~\eqref{StressMax}. Using Eq.~\eqref{MagRelations}, we rewrite the total stress as
\begin{equation}
	\begin{aligned}
		\bm{\sigma} &= J^{-1} \frac{\partial \Omega^\text{e}}{\partial \bm F} \bm{F}^\text{T}
	- \mu^{-1} (\mathbb{b} \otimes \mathbb{b}^\text{r} + \mathbb{b}^\text{r} \otimes \mathbb{b})  \\
	&\quad + \mu^{-1} (\mathbb{b}^\text{r} \cdot\mathbb{b})\bm{I} + \mu^{-1} \big(\mathbb{b}\otimes\mathbb{b} - \tfrac12 |\mathbb{b}|^2\bm{I} \big) ,
	\end{aligned}
	\label{StressHMAE}
\end{equation}
which can be obtained directly from Eq.~\eqref{HMAE}\textsubscript{1} by evaluation of $J^{-1} (\Omega_{\bm F}) {\bm F}^\text{T}$ with $\mathbb{B}^\text{r}$ constant.
Consistently with Eq.~\eqref{Equil}\textsubscript{2}, we observe that $\bm \sigma$ is a symmetric tensor \citep{dorfmann24}.

It is worth noting that the remnant magnetic induction $\mathbb{B}^\text{r}$ simply acts as a parameter which is not constitutively determined. Therefore, the Clausius--Duhem inequality \eqref{Dissipation} does not depend on this quantity or on its evolution. In other words, the present theory might be labelled \emph{pseudo}-magneto-elastic. Further discussions about such theories can be found in the studies by \citet{danas24} and by \citet{gebhart25}, where formulations with internal variables of state are also presented. In this case, the evolution of the internal variables is constitutively determined via the Clausius--Duhem inequality to enforce a dissipative material behaviour, see also the review by \citet{lucarini22}. As proposed by \citet{morro24}, one might also add the variable $\mathbb M$ to the variables of state $\lbrace\bm{F}, \mathbb{H}\rbrace$ to describe a hysteresis effect. In this case, the magnetisation is deduced from constitutive equations, which need to comply with thermodynamic restrictions expressed in terms of the potential energy $\Omega^*$ introduced in Eq.~\eqref{Legendre}.

\begin{remark}
	In the derivation of the constitutive law \eqref{StressHMAE} from Eq.~\eqref{HMAE}\textsubscript{1}, the remnant magnetic induction $\mathbb{B}^\text{r}$ in the reference configuration was treated like a constant parameter, in a similar fashion to \citet{zhao19}. If this derivation was performed using Eq.~\eqref{HMAE}\textsubscript{2} with a constant remnant magnetisation $\mathbb{M}^\text{r}$, then we would find
	\begin{equation}
		\bm{\sigma} = J^{-1} \frac{\partial \Omega^\text{e}}{\partial \bm F} \bm{F}^\text{T} + \mu^{-1} \big(\mathbb{b}\otimes\mathbb{b} - \tfrac12 |\mathbb{b}|^2\bm{I} \big) ,
	\end{equation}
	instead of \eqref{StressHMAE}. Again, the total stress $\bm\sigma$ is a symmetric tensor, and the linear law \eqref{MagLinear} still holds. However, with this choice, $\bm\sigma$ turns out to be independent on $\mathbb{M}^\text{r}$. In addition, note that the present definition $\mathbb{M} = \bm{F}^\text{T}\mathbb{m}$ of the magnetisation in the reference configuration \eqref{Magnetisation} is not the only possibility. Expressions corresponding to the alternative definition $\mathbb{M} = J\bm{F}^{-1}\mathbb{m}$ are provided by \citet{dorfmann24}.
\end{remark}

\begin{remark}
	Here we have followed the same steps as in \citet{zhang22}, up to Eq.~(23) therein. Based on the observation that the remnant magnetisation is stretch-independent, a different definition for $\mathbb{M}^\text{r} \neq \bm{F}^\text{T}\mathbb{m}^\text{r}$ is used in \cite{zhang22}. In their case, the remnant magnetisation satisfies $\mathbb{M}^\text{r} = \bm{R}^\text{T}\mathbb{m}^\text{r}$, where
\begin{equation}
	\bm{F} = \bm{R}\bm{U} = \bm{V} \bm{R},
\end{equation}	
defines the unique right polar decomposition of the deformation gradient ($\bm{F} = \bm{R}\bm{U}$) as well as its unique left polar decomposition ($\bm{F} = \bm{V} \bm{R}$). Here, the rotation tensor $\bm R$ is orthogonal, and the stretch tensors $\bm{U}$, $\bm V$ are positive definite and symmetric \citep{holzapfel00}. Using these definitions, it follows that $\mathbb{M}^\text{r} = (\mu_0 J)^{-1} \bm{R}^\text{T} \bm{F}\mathbb{B}^\text{r}$, which can be used to rewrite Eq.~\eqref{HMAE}\textsubscript{1} in the same form as Eq.~(25) of the discussed study. In Eq.~(26) of \cite{zhang22}, the total stress tensor $\bm{\sigma} = J^{-1} (\Omega_{\bm F}) {\bm F}^\text{T}$ is not symmetric, which results from treating $\mathbb{m}^\text{r} = \bm{R} \mathbb{M}^\text{r}$ like a constant when evaluating $\Omega_{\bm F}$. If instead $\mathbb{M}^\text{r}$ was treated like a constant, then the calculation of $\Omega_{\bm F}$ would require an examination of the functional dependence of $\bm R$ with respect to $\bm F$ (cf. \citet{rosati99}).
\end{remark}

\paragraph{Incompressibility}

In incompressible materials, the volume dilation \eqref{VolumeChange} is prescribed, i.e., $J \equiv 1$ is enforced, and the mass density $\rho = \rho_0$ does not change over time. Since volume change is not allowed, the constraint of incompressibility simplifies the expression of the magneto-elastic energy \eqref{HMAE} which does not depend on $J$. In general, the strain energy function $\Omega^\text{e}$ can be expressed as a function of two scalar invariants $I_1$ and $I_2$,
\begin{equation}
	\Omega^\text{e} = W(I_1, I_2) , \quad I_1 = \text{tr}\, \bm{C}, \quad I_2 = \text{tr}\, \bm{C}^{-1} .
	\label{ElastEnergyGen}
\end{equation}
The linear law \eqref{MagLinear} for the magnetic field still holds. The total stress $\bm{\sigma} = (\Omega_{\bm F}) \bm{F}^\text{T}$ deduced from \eqref{StressHMAE} is expressed as follows,
\begin{equation}
	\begin{aligned}
	\bm{\sigma} &= \frac{G}{1 + \upsilon} \left( \bm{B} - \upsilon \bm{B}^{-1} \right) \\ 
	&\quad  + \mu^{-1} (\mathbb{b}\otimes\mathbb{b} - \mathbb{b} \otimes \mathbb{b}^\text{r} - \mathbb{b}^\text{r} \otimes \mathbb{b})- p\bm{I} ,
	\end{aligned}
	\label{StressIncomp}
\end{equation}
where $\bm{B} = \bm{F}\bm{F}^\text{T}$ is the left Cauchy--Green strain tensor, and $p$ is an arbitrary Lagrange multiplier to enforce the incompressibility constraint ($J \equiv 1$). The coefficient $G = 2(W_1+W_2)$ represents the generalised shear modulus of an undeformed material (see \citet{destrade08} for a generalisation to deformed media), where the functions $W_i = \partial W/\partial I_i$ are partial derivatives of $W$, and $\upsilon = W_2/W_1$. As mentioned in the previous remark, the term proportional to $\mathbb{b}\otimes \mathbb{b}^\text{r}$ above is missing in Eq.~(26) of \citet{zhang22}, where the stress tensor is not symmetric.

In what follows, we consider magneto-elastic laminates whose layers are described by the generalised neo-Hookean strain energy function
\begin{equation}
	\Omega^\text{e} = W(I_1) ,
	\label{ElastEnergy}
\end{equation}
for which $\upsilon = 0$, and $G = 2W_1$ is a function of $I_1$ only. This restriction leads to a number of useful analytical results and simplifications, see for instance \citet{boulanger92} for a discussion of elastic wave propagation in deformed Mooney--Rivlin solids. The study of the general case \eqref{ElastEnergyGen} could become the scope of future investigations.

\begin{figure*}
	\centering
	
	\begin{minipage}{0.5\textwidth}
		\centering
		(a)
	\end{minipage} \hfill
	\begin{minipage}{0.49\textwidth}
		\centering
		(b)
	\end{minipage}
	
	\begin{minipage}{0.5\textwidth}
		\centering
		\includegraphics{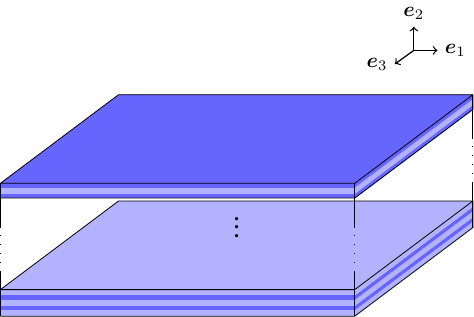}
	\end{minipage} \hfill
	\begin{minipage}{0.49\textwidth}
		\centering
		\includegraphics{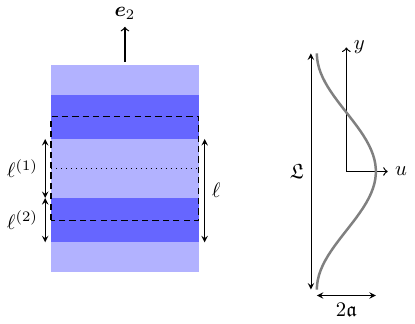}
	\end{minipage}
	
	\caption{Periodic laminate in the deformed configuration. (a) Global view, and (b) zoom on the vertical cross-section. The region marked by dashed lines indicates the periodically repeated unit cell. The parameters of the schematic waveform shown on the right are the amplitude $\mathfrak{a}$ and the wavelength $\mathfrak{L}$. \label{fig:Struct}}
\end{figure*}

\section{Magnetoelastic laminates}\label{sec:Laminate}

Consider a periodic laminate consisting of two isotropic and incompressible alternating layers with volume fractions $\nu^{(1)}$ and $\nu^{(2)} = 1-\nu^{(1)}$. The lamination direction is along the $y$-axis, which is oriented by the unit vector $\bm{e}_2$. Denoting the spatial period of the undeformed laminate as $L$, the layer thicknesses equal $L^{(\alpha)} = \nu^{(\alpha)}L$ in the undeformed laminate, where the exponents $\alpha \in \lbrace 1, 2\rbrace$ in parentheses mark each layer. The composite sample is a rectangular block of large but finite size which is surrounded by vacuum (Figure~\ref{fig:Struct}a). An external magnetic field is applied in the lamination direction, and no other load is applied on the boundaries. The number of alternating layers is very large but finite, and with the same number of layers for each material. We consider an idealisation of the periodic microstructure shown in Fig.~\ref{fig:Struct}, where we restrict our attention to describing the motion at a significant distance from the edges of the sample. This way, the mechanical and magnetic fields are homogeneous in each layer of the laminate and are determined by the appropriate jump conditions.

\subsection{Static magneto-deformation}

Similarly to \citet{zhang22}, we consider the case where the remnant magnetisation is oriented along the lamination direction, i.e.,
\begin{equation}
	\mathbb{m}^\text{r($\alpha$)} = m^\text{r($\alpha$)}\bm{e}_2, \quad \mathbb{b}^\text{r($\alpha$)} = b^\text{r($\alpha$)}\bm{e}_2,
	\label{MagStatic}
\end{equation}
for each layer, where $b^\text{r($\alpha$)} = \mu_0 m^\text{r($\alpha$)}$ is a constant (i.e., the remnant magnetisation does not vary with respect to the spatial coordinates within one phase, as we shall see in the upcoming paragraphs). Both layers are described by the constitutive law \eqref{StressIncomp} and the energy function \eqref{HMAE}-\eqref{ElastEnergy}, with the elastic energy $W^{(\alpha)}$, the magnetic permeability $\mu^{(\alpha)}$, the generalised shear modulus $G^{(\alpha)}$, and the Lagrange multiplier $p^{(\alpha)}$ to enforce the incompressibility constraint.

A uniform and permanent magnetic induction field $\mathbb{b} = b \bm{e}_2$ is applied to the surrounding vacuum, and the magnetic induction in the layered material is sought in the form $\mathbb{b}^{(\alpha)} = b^{(\alpha)} \bm{e}_2$. The magnetic boundary conditions at the inner layers' interfaces require that the magnetic induction is continuous across them, and Gauss' law of magnetostatics implies that $b^{(\alpha)}$ is independent of $y$ within each layer. Finally, the magnetic boundary conditions at the top and bottom surfaces of the sample require that $b^{(\alpha)} = b$ is a constant. According to \eqref{MagLinear}, the magnetic field $\mathbb{h}^{(\alpha)} = {h}^{(\alpha)} \bm{e}_2$ in each layer is also oriented along $\bm{e}_2$, and we have $h^{(\alpha)} = (b - b^\text{r($\alpha$)})/\mu^{(\alpha)}$. Obviously, the above identities hold only far from the lateral boundaries since $\mathbb{h}^{(\alpha)} \times \bm{e}_1$ is not continuous across the boundaries with unit normal $\bm{e}_1$.

For this configuration, we seek the deformation gradient tensor in each layer in the following form,
\begin{equation}
 	\bm{F}^{(\alpha)} =  \lambda^{(\alpha)} \bm{e}_2 \otimes \bm{e}_2 + (\lambda^{(\alpha)})^{-1/2} \left(\bm{I} - \bm{e}_2 \otimes \bm{e}_2\right) ,
 	\label{FStretch}
\end{equation}
where the stretch $\lambda^{(\alpha)} > 0$ along the $y$-axis in the layer $\alpha$ is assumed constant (i.e., independent of space and time). For such an uniaxial tensile motion described by the mapping $\bm{x} = \bm{F}^{(\alpha)}\bm{X}$, the incompressibility constraint is naturally satisfied at every point. We note that when the material is stretched in the lamination direction $(\lambda^{(\alpha)} > 1)$, it is simultaneously compressed in the transverse direction, i.e., $(\lambda^{(\alpha)})^{-1/2} < 1$. Furthermore, the remnant magnetic induction $\mathbb{b}^\text{r($\alpha$)} = \bm{F}^{(\alpha)}\mathbb{B}^\text{r($\alpha$)}$ is a constant vector within one phase provided $\mathbb{B}^\text{r($\alpha$)}$ is constant, which is coherent with earlier assumptions \eqref{MagStatic}.

For a perfectly bonded laminate, the continuity of the displacement field $\bm{x}-\bm{X}$ at the interface between two adjacent layers requires that
\begin{equation}
	(\bm{F}^{(1)} - \bm{I})\bm{X} = (\bm{F}^{(2)} - \bm{I})\bm{X} ,
\end{equation}
for every $\bm X$ belonging to the interface. Thus, the tensor $\bm{F}^{(1)} - \bm{F}^{(2)}$ must be singular, and we conclude that the stretching ratio is uniform:
\begin{equation}
	\lambda^{(1)} = \lambda^{(2)} = \lambda .
	\label{StretchConst}
\end{equation}
The stress tensor $\bm{\sigma}^{(\alpha)}$ in each layer is deduced from Eq.~\eqref{StressIncomp} with $\upsilon = 0$, the deformation gradient tensor $\bm{F}^{(\alpha)}$, the Lagrange multiplier $p^{(\alpha)}$, the magnetic permeability $\mu^{(\alpha)}$, and the generalised shear modulus $G^{(\alpha)}$. The latter is function of the invariant $I_1 = \bm{F}^{(\alpha)} : \bm{F}^{(\alpha)}$ (hence, of the stretch $\lambda$), in general.

Based on the above results, we are now able to provide a description of the deformed laminate. If we introduce the thicknesses $\ell^{(\alpha)} = \lambda L^{(\alpha)}$ of each layer as well as the spatial period $\ell = \lambda L$ of the deformed laminate, then we have the relationship $\ell^{(\alpha)} = \nu^{(\alpha)} \ell$. We have represented these quantities in Figure~\ref{fig:Struct}b, where the periodic unit cell is marked by dashed lines. In a next step, we use the mechanical equilibrium equations to express the relationship between the stretch $\lambda$ and the external magnetic induction along with the remnant magnetic induction.

The equilibrium condition \eqref{Equil} with $\bm{a} = \bm{0}$ implies that $p^{(\alpha)}$ is constant in each layer. In addition, since the sample is surrounded by vacuum, we note that the normal tractions $\bm{t}^{(\alpha)} = \bm{\sigma}^{(\alpha)} \bm{e}_2$ with $\alpha = 1,2$ on the upper and lower external boundaries are both equal to the normal traction ${\bm t}^\text{m} = \bm{\sigma}^\text{m} \bm{e}_2$ deduced from the Maxwell stress \eqref{StressMax}, which provides us the expression of the Lagrange multipliers $p^{(\alpha)}$ in terms of the stretch and magnetic induction. We note in passing that the normal tractions $\bm{t}^{(\alpha)}$ in the lamination direction are naturally continuous across the interfaces between two adjacent layers. Finally, we introduce the traction $\bm{s}^\text{m} = \bm{\sigma}^\text{m} \bm{e}_1$ applied by the surrounding vacuum to the lateral faces, which must be equal to the average lateral traction $\bar{\bm s} = \nu^{(1)}\bm{s}^{(1)} + \nu^{(2)}\bm{s}^{(2)}$ inside the composite material, where $\bm{s}^{(\alpha)} = \bm{\sigma}^{(\alpha)} \bm{e}_1$. By eliminating the Lagrange multipliers $p^{(\alpha)}$ from this equation, we arrive at
\begin{equation}
	 \mu_0\overline{G} \left(\lambda^{2}-\lambda^{-1}\right) = \left(1 - \mu_0/\breve \mu\right) b^2 + 2 b \check{b}^\text{r} \mu_0 / \breve \mu ,
	 \label{StretchMag}
\end{equation}
where
\begin{equation}
	\breve\mu = \left(\frac{\nu^{(1)}}{\mu^{(1)}} + \frac{\nu^{(2)}}{\mu^{(2)}}\right)^{-1} , \quad
	\frac{\check{b}^\text{r}}{\breve\mu} = \frac{\nu^{(1)} b^\text{r(1)}}{\mu^{(1)}} + \frac{\nu^{(2)} b^\text{r(2)}}{\mu^{(2)}} .
	\label{StretchCoeff}
\end{equation}
In general, the average shear modulus
\begin{equation}
	\overline{G} = \nu^{(1)} G^{(1)} + \nu^{(2)} G^{(2)}
	\label{AverageG}
\end{equation}
is a function of the invariant $I_1 = \lambda^2 + 2/\lambda$.
Formally, this result matches Eq.~(43) of \citet{zhang22}, up to the coefficient two in the right-hand side of \eqref{StretchMag} which is a direct consequence of the symmetric form of the stress tensor \eqref{StressIncomp}. Consequences of the relationship \eqref{StretchMag} are presented in Section~\ref{sec:Tune}.

A comparable situation is that of a laminated slab in vacuum which is infinite along one direction transverse to the lamination direction ($\bm{e}_2$), say along $\bm{e}_3$. The slab is semi-infinite in the other direction, so that there is one free surface with unit normal $\bm{e}_1$ and two free surfaces with unit normal $\bm{e}_2$ corresponding to the upper and lower faces. By imposing the tractions due to the Maxwell stress \eqref{StressMax} on these surfaces and by following the same steps as above, the same relationship \eqref{StretchMag} between the stretch and the magnetic induction can be obtained, at a significant distance from the lateral boundary with unit normal $\bm{e}_1$.

Another comparable situation is that of a laminated block (or of a laminated semi-infinite slab) which has only one free boundary with unit normal $\bm{e}_2$, say the upper face. The other boundary of this type, on the underside, is not free but clamped (i.e., the displacement field deduced from the deformation \eqref{FStretch} is imposed){\,---\,}or no such boundary on the lower face is considered (i.e., the medium is assumed semi-infinite in the lamination direction). In either case, the free surfaces are subjected to the tractions due to the Maxwell stress \eqref{StressMax}, and the normal tractions must be continuous across the inner interfaces. By following the same steps as earlier, the relationship \eqref{StretchMag} between the stretch and the magnetic induction is recovered (at a sufficient distance from the lateral boundaries).

An effective energy function can be deduced from the above expressions by spatial averaging. Here, the magnetoelastic energy \eqref{HMAE}-\eqref{ElastEnergy} in each layer reads
\begin{equation}
	\Omega^{(\alpha)} = W^{(\alpha)}(I_1) - b b^\text{r$(\alpha)$}/\mu^{(\alpha)} + \tfrac12 b^2/\mu^{(\alpha)}  ,
\end{equation}
with $I_1 = \lambda^2 + 2/\lambda$. The spatial average of this expression takes the form
\begin{equation}
	\begin{aligned}
	\overline{\Omega} &= \nu^{(1)}\Omega^{(1)} + \nu^{(2)}\Omega^{(2)} \\
	&= \overline{W}(I_1) - \breve\mu^{-1} \mathbb{b} \cdot \check{\mathbb{b}}^\text{r} + \tfrac12 \breve\mu^{-1} \mathbb{b}^2 ,
	\end{aligned}
\end{equation}
where $\overline{W} = \nu^{(1)}W^{(1)} + \nu^{(2)}W^{(2)}$. The effective remnant magnetisation reads
\begin{equation}
	\frac{\check{\mathbb{b}}^\text{r}}{\breve\mu}  = \frac{\nu^{(1)} \mathbb{b}^\text{r(1)}}{\mu^{(1)}} + \frac{\nu^{(2)} \mathbb{b}^\text{r(2)}}{\mu^{(2)}} ,
\end{equation}
which leads to a similar energy function as the function $\Omega$ introduced in Eqs.~\eqref{HMAE}, in the incompressible limit ($J=1$). It can be shown that a homogeneous and incompressible sample described by the energy function $\overline{\Omega}$ and subjected to the same magneto-deformation as the layered material produces the same relationship \eqref{StretchMag} between the stretch and the magnetic field. Therefore, the above strain energy function corresponds to a homogenised material theory for a sample subjected to special mechanical and magnetic conditions. In general, a suitable effective magneto-elastic energy should depend on the lamination direction, $\bm{n}$. This observation is exploited in Section~\ref{sec:Homogen} where we focus on the elastic part of the energy, $\overline{W}$.

\subsection{Nonlinear shear waves}

Based on the above static magneto-deformation, we now consider a dynamic problem which consists of a time-dependent perturbation of the equilibrium configuration described in the previous subsection. As noted therein, laminates that obey the law \eqref{StretchMag} can have various geometries, such as rectangular blocks with six faces (Fig.~\ref{fig:Struct}a), or semi-infinite bodies with at least one free boundary along $\bm{e}_2$ and another one perpendicular to $\bm{e}_2$. Hence, we can place ourselves at a large distance from the lateral boundaries where the static magneto-deformation is known, and there, consider plane shear waves that propagate along the $y$-axis (orientated by $\bm{e}_2$) with a polarisation along the $x$-axis (orientated by $\bm{e}_1$). In addition, we assume that the sample is large enough to be considered of infinite size along $y$ for the present one-dimensional wave propagation study.

We consider finite amplitude shear waves described by the total deformation
\begin{equation}
	x = \lambda^{-1/2} X + u, \quad y = \lambda Y, \quad z = \lambda^{-1/2} Z ,
	\label{MotShear}
\end{equation}
in each layer.
The displacement field $u$ along the $x$-axis is function of the vertical coordinate $Y$ and of the time $t$ only, so that the deformation gradient tensor \eqref{FStretch}-\eqref{StretchConst} becomes
\begin{equation}
	\bm{F}^{(\alpha)} = \begin{bmatrix}
		\lambda^{-1/2} & u_Y & 0 \\
		0 & \lambda & 0 \\
		0 & 0 & \lambda^{-1/2}
	\end{bmatrix}.
	\label{FShear}
\end{equation}
Similarly, the magnetic field $\mathbb{h}^{(\alpha)}$ and the magnetic induction $\mathbb{b}^{(\alpha)}$ are functions of the vertical coordinate $Y$ and of the time $t$ only. The remnant magnetic induction $\mathbb{b}^\text{r($\alpha$)}$ evaluated in the static case is assumed to be unaffected by the wave propagation process (i.e., fluctuations in this variable are neglected), so that $\mathbb{b}^\text{r($\alpha$)} = {b}^\text{r($\alpha$)} \bm{e}_2$ can be treated like a constant vector. This assumption made in \citet{zhang22} is a point that could be improved in the future{\,---\,}the picture is slightly different in \citet{alam23} where $\mathbb{B}^\text{r($\alpha$)}$ and $\mathbb{b}$ are treated like constants.

Maxwell's equations \eqref{Maxwell} together with the relationship \eqref{MagLinear} where the remnant magnetic induction $\mathbb{b}^\text{r($\alpha$)}$ and magnetic permeability $\mu^{(\alpha)}$ are constants require that $\mathbb{h}^{(\alpha)}$ and $\mathbb{b}^{(\alpha)}$ are both divergence-free and curl-free in each layer. Exploiting invariance with respect to the $x$ and $z$ coordinates, we find that $\mathbb{h}^{(\alpha)}$ and $\mathbb{b}^{(\alpha)}$ must be spatially uniform within layers.
Using the constitutive law \eqref{StressIncomp}-\eqref{ElastEnergy} and invariance with respect to the $x$ and $z$ coordinates, the equation of motion \eqref{Equil} reduces to
\begin{equation}
	(\lambda^2 G^{(\alpha)} u_{y})_{y} = \rho^{(\alpha)} u_{tt} ,
	\quad
	(\lambda^2 G^{(\alpha)} - p^{(\alpha)})_y = 0 ,
	\label{WaveReal}
\end{equation}
in each layer, where indices indicate partial differentiation with respect to space and time. The Lagrange multipliers $p^{(\alpha)}$ are functions of $y$ and $t$, and we have used the relationship $(\cdot)_Y = \lambda (\cdot)_y$ between spatial differentiation in the deformed and undeformed configurations.

The vertical component of the magnetic induction $\mathbb{b}^{(\alpha)}$ and the tangential components of the magnetic field $\mathbb{h}^{(\alpha)}$ are required to be continuous across the layer interfaces. A specific solution that satisfies all these requirements takes the same form as in the static case. In fact, it suffices to set $\mathbb{b}^{(\alpha)} = b \bm{e}_2$ and $\mathbb{h}^{(\alpha)} = h^{(\alpha)} \bm{e}_2$ where $h^{(\alpha)} = (b-b^\text{r($\alpha$)})/\mu^{(\alpha)}$. In addition, the displacement field $u$ is required to be continuous across the layer interfaces. A similar continuity requirement holds for the normal tractions $\bm{t}^{(\alpha)} = \bm{\sigma}^{(\alpha)} \bm{e}_2$, which leads to the continuity of
\begin{equation}
	\lambda^2 G^{(\alpha)} u_{y} ,\quad
	\lambda^2 G^{(\alpha)} - p^{(\alpha)} + ( b^2 - 2b{b}^\text{r($\alpha$)}) /\mu^{(\alpha)} ,
\end{equation}
across layer interfaces, where the first quantity $\lambda^2 G^{(\alpha)} u_{y}$ is the shear stress of Eq.~\eqref{WaveReal}\textsubscript{1}. The continuity of the second quantity above allows to determine the pressure jump $p^{(2)}-p^{(1)}$ across layer interfaces.

Here, the generalised shear modulus $G^{(\alpha)}$ is function of
\begin{equation}
	I_1 = \mathcal{I}_1 + \lambda^2 u_y^2, \quad \mathcal{I}_1 = \lambda^2 + 2/ \lambda ,
	\label{WaveI1}
\end{equation}
and the shear strain $u_y$ is governed by a nonlinear wave equation deduced from \eqref{WaveReal}\textsubscript{1} by differentiation with respect to $y$.
We assume that $G^{(\alpha)}$ is smooth enough to be expanded as a Taylor series in the vicinity of $I_1 = \mathcal{I}_1$, so that we can write
\begin{equation}
	\begin{aligned}
	\lambda^2 G^{(\alpha)} &\simeq \lambda^2 G^{(\alpha)}|_{I_1 = \mathcal{I}_1} + \lambda^4 G^{(\alpha)\prime}|_{I_1 = \mathcal{I}_1}  u_y^2\\
	&= g^{(\alpha)} + \tfrac{1}{3} h^{(\alpha)} u_y^2 ,
	\end{aligned}
	\label{ShearNL}
\end{equation}
where $G^{(\alpha)\prime}$ is the derivative of $G^{(\alpha)}$ with respect to $I_1$, from which we identify the coefficients $g^{(\alpha)}$ and $h^{(\alpha)}$. The above expression of $\lambda^2 G^{(\alpha)}$ is then injected in the wave equation \eqref{WaveReal}\textsubscript{1}, which produces the wave equation \eqref{WaveAlpha}. In the limit of small wave amplitudes, $|h^{(\alpha)}u_y^2| \ll g^{(\alpha)}$, we recover the same wave equation in each layer as \citet[Eq.~(48)]{zhang22}, in which the squared wave speed $(\mathfrak{c}^{(\alpha)})^2$ is equal to $\lambda^{2} G^{(\alpha)}/\rho^{(\alpha)}$.

Under our modelling assumptions, we note that the equation \eqref{WaveAlpha} governing the propagation of a plane shear wave along the direction of a permanent magnetic induction is identical to the one obtained for magnetically inert media \cite{destrade08,chocka20}. Thus, we could ignore the magneto-elastic origin of the static deformation \eqref{FStretch}. The present wave propagation problem is much simpler than that obtained for some soft-magnetic material models, where the dynamic shear wave field is coupled to a time-dependent component of the magnetic induction, in general \citep{destrade11}.

\begin{table*}
	\caption{ Strain energy function $W^{(\alpha)}$, generalised shear modulus $G^{(\alpha)} = 2 W^{(\alpha)}_1$ and its derivative $G^{(\alpha)\prime}$ for a variety of generalised neo-Hookean materials \eqref{ElastEnergy}. We provide also the relationship to the coefficients $g^{(\alpha)} = \lambda^2 G^{(\alpha)}|_{I_1 = \mathcal{I}_1}$ and $h^{(\alpha)} = 3\lambda^4 G^{(\alpha)\prime}|_{I_1 = \mathcal{I}_1}$ which govern shear wave propagation \eqref{WaveAlpha}. Here, we have introduced the coefficient $\underline{\beta}^{(\alpha)}$ which is the value of ${\beta}^{(\alpha)}$ obtained for the Fung-Demiray theory, for known values of $g^{(\alpha)}$ and $h^{(\alpha)}$. The letter `$\text{e}$' stands for Euler's number, $\exp(1)$. \label{tab:Elast}}
	\vspace{0.5em}
	\centering
	{\renewcommand{\arraystretch}{1}
	\small
	\begin{tabular}{cccc}
    \toprule 
      & Yeoh & Fung-Demiray & Gent \\
     \midrule
     $W^{(\alpha)}$ & $\displaystyle \frac{\mathcal{G}^{(\alpha)}}{2} \left( (I_1 - 3) + \tfrac12 \beta^{(\alpha)} (I_1 - 3)^2 \right)$ & $\displaystyle \frac{\mathcal{G}^{(\alpha)}}{2 \beta^{(\alpha)}} \left(\text{e}^{\beta^{(\alpha)} (I_1-3)}- 1\right)$ & $\displaystyle -\frac{\mathcal{G}^{(\alpha)}}{2 \beta^{(\alpha)}} \ln\left(1 - \beta^{(\alpha)} (I_1-3)\right)$ \\[0.5em]
     $G^{(\alpha)}$ & $\mathcal{G}^{(\alpha)} \left( 1 + \beta^{(\alpha)} (I_1 - 3) \right)$ & $\mathcal{G}^{(\alpha)} \text{e}^{\beta^{(\alpha)} (I_1-3)}$ & $\mathcal{G}^{(\alpha)} \left(1 - \beta^{(\alpha)} (I_1-3)\right)^{-1}$ \\[0.5em]
     $G^{(\alpha)\prime}$ & $\mathcal{G}^{(\alpha)} \beta^{(\alpha)}$ & $\mathcal{G}^{(\alpha)} \beta^{(\alpha)} \text{e}^{\beta^{(\alpha)} (I_1-3)}$ & $\mathcal{G}^{(\alpha)}\beta^{(\alpha)} \left(1 - \beta^{(\alpha)} (I_1-3)\right)^{-2}$ \\
     \midrule
     $\mathcal{G}^{(\alpha)}$ & $\displaystyle \frac{g^{(\alpha)}}{\lambda^{2}} \left( 1 - \underline{\beta}^{(\alpha)} (\mathcal{I}_1-3) \right)$ & $\displaystyle \frac{g^{(\alpha)}}{\lambda^{2}} \text{e}^{- \underline{\beta}^{(\alpha)} (\mathcal{I}_1-3) }$ & $\displaystyle \frac{g^{(\alpha)}}{\lambda^{2}} \left( 1 + \underline{\beta}^{(\alpha)} (\mathcal{I}_1-3) \right)^{-1}$ \\[0.5em]
     $\beta^{(\alpha)}$ & $\displaystyle \left( 1/\underline{\beta}^{(\alpha)} - (\mathcal{I}_1-3) \right)^{-1}$ & $\displaystyle \underline{\beta}^{(\alpha)} = \frac{h^{(\alpha)}}{3\lambda^2 g^{(\alpha)}}$ & $\displaystyle \left( 1/\underline{\beta}^{(\alpha)} + (\mathcal{I}_1-3) \right)^{-1}$ \\
    \bottomrule 
    \end{tabular}}
\end{table*}

For the sake of illustration we consider a variety of suitable strain energy functions of generalised neo-Hookean type \eqref{ElastEnergy}, namely the two-term Yeoh theory, the Fung-Demiray theory, and the Gent theory, see expressions in Table~\ref{tab:Elast}. For all these constitutive laws, the parameters $\mathcal{G}^{(\alpha)}$ and $\beta^{(\alpha)}$ are usually assumed positive, and the neo-Hookean theory is recovered in the limit $\beta^{(\alpha)} \to 0$. We then calculate the generalised shear moduli $G^{(\alpha)} = 2 W^{(\alpha)}_1$, as well as their derivative $G^{(\alpha)\prime}$ where the prime indicates differentiation with respect to $I_1$. The coefficients of Eq.~\eqref{ShearNL} are determined uniquely based on these expressions. Inversely, we can calculate the parameters $\mathcal{G}^{(\alpha)}$ and $\beta^{(\alpha)}$ for known values of $g^{(\alpha)}$ and $h^{(\alpha)}$. The relevant expressions are also provided in Table~\ref{tab:Elast}.

\begin{figure}
	\centering
	\includegraphics{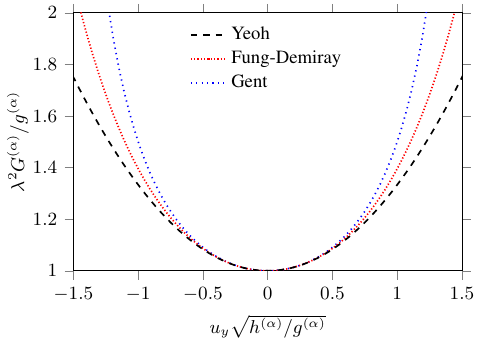}
	\caption{Generalised shear stiffness $\lambda^2 G^{(\alpha)}$ in terms of a normalised shear strain, where $u_y$ is the shearing ratio. The quantity displayed is evaluated at $I_1 = \mathcal{I}_1$ and then normalised, see Eq.~\eqref{ShearNL}; $\lambda$ is the stretch ratio of an uniaxial deformation along $y$. \label{fig:Elast}}
\end{figure}

To highlight the similarities and differences between the above constitutive laws, we consider a homogeneous and magnetically inert elastic solid described by the strain energy functions of Table~\ref{tab:Elast}. The solid is subjected to the deformation \eqref{FShear} of stretch ratio $\lambda$, and the stiffness coefficient $\lambda^2 G^{(\alpha)}$ of Eq.~\eqref{WaveReal}\textsubscript{1} is calculated. Evaluating this quantity at the value of $I_1$ specified in Eq.~\eqref{WaveI1} and using the expression of $\beta^{(\alpha)}$ in Table~\ref{tab:Elast}, we find that the ratio $\lambda^2 G^{(\alpha)}/g^{(\alpha)}$ equals $1 + \frac13 r^2$ for the Yeoh theory, where $r^2$ is equal to the quantity $u_y^2 h^{(\alpha)} / g^{(\alpha)}$. In contrast, we find that $\lambda^2 G^{(\alpha)}/g^{(\alpha)}$  equals $\exp(\frac13 r^2)$ for the Fung-Demiray theory, and $(1-\frac13 r^2)^{-1}$ for the Gent theory.

The above functions of $r$ are represented in Fig.~\ref{fig:Elast}. Based on these curves, we note that the Yeoh, Fung-Demiray and Gent models agree very well for any axial stretch and over a large range of shear strains, which implies that these theories can be used interchangeably within this range. In fact, for small values of $r^2 \ll 1$, all these constitutive laws satisfy the asymptotic equivalence $\lambda^2 G^{(\alpha)}/g^{(\alpha)} \sim 1 + \frac13 r^2$, although terms of order $r^4$ differ. In the case of incompressible hard-magnetic materials, this observation could also be restated in terms of restrictions on the magnetic fields $b$ and $\check{b}^\text{r}$ instead of $r$, see Eq.~\eqref{StretchMag} and Table~\ref{tab:Elast}.

\begin{remark}
The problem of shear wave propagation with an arbitrary wave polarisation in the $xz$-plane can be studied in a similar fashion. Introducing the displacement $w(Y,t)$ such that $z = \lambda^{-1/2} Z + w$ in Eq.~\eqref{MotShear} produces two coupled wave equations instead of Eq.~\eqref{WaveReal}\textsubscript{1}. In fact, the above steps lead to the pair of wave equations
\begin{equation}
(\lambda^2 G^{(\alpha)} u_{y})_{y} = \rho^{(\alpha)} u_{tt} , \quad (\lambda^2 G^{(\alpha)} w_{y})_{y} = \rho^{(\alpha)} w_{tt} ,
\end{equation}
where $I_1$ depends on the strain variables $u_y$ and $w_y$, and so does the generalised shear modulus $G^{(\alpha)}$ as well. These equations can be rewritten in a condensed form by introducing the complex variable $\mathscr{U} = u + \text{i} w$,
\begin{equation}
	(\lambda^2 G^{(\alpha)} \mathscr{U}_{y})_{y} = \rho^{(\alpha)} \mathscr{U}_{tt} ,
\end{equation}
see for instance \citet{destrade08}.
Here, the quantity $\lambda^2 G^{(\alpha)} \mathscr{U}_{y}$ is required to be continuous across layer interfaces, and $G^{(\alpha)}$ is function of $\mathcal{I}_1 + \lambda^2|\mathscr{U}_y|^2$. Therefore, this problem is formally analogous to the one considered earlier \eqref{WaveReal}-\eqref{WaveI1}, where we have assumed that $w$ is identically equal to zero.
\end{remark}

\section{Homogenisation theory}\label{sec:Homogen}

\subsection{Two-scale asymptotic analysis}

To understand the effective dynamic response of the pre-deformed laminate, we develop a homogenisation theory similar to \citet{andrianov13}, where the case of nonlinear compression waves was addressed, for which the stress is a quadratic polynomial expression of the strain (instead of the present cubic nonlinearity, see previous section). To this aim, we expand the equation of motion \eqref{WaveReal}\textsubscript{1}-\eqref{ShearNL} using the chain rule, and rewrite it in terms of dimensionless coordinates,
\begin{equation}
	\begin{aligned}
	& \left(\text{g}^{(\alpha)} + \text{h}^{(\alpha)} \delta^2 \text{u}_\text{y}^2\right) \text{u}_\text{yy} = \varrho^{(\alpha)} \text{u}_\text{tt} , \\
	&\text{u} = u/\mathfrak{a}, \quad \text{y} = y/\mathfrak{L}, \quad \text{t} = \mathfrak{c} t/\mathfrak{L} , \quad \delta = \mathfrak{a}/\mathfrak{L},\\
	& \text{g}^{(\alpha)} = g^{(\alpha)}/ \langle g \rangle , \quad \text{h}^{(\alpha)} = h^{(\alpha)}/ \langle g \rangle , \quad \varrho^{(\alpha)} = \rho^{(\alpha)}/\langle\rho\rangle ,
	\end{aligned}
	\label{WaveFinal}
\end{equation}
where the positive parameters $\mathfrak{a}$, $\mathfrak{L}$, $\mathfrak{c}$ denote the effective amplitude, the spatial length, and the velocity of a wave, respectively (see Fig.~\ref{fig:Struct}b for a schematic depiction of the wave amplitude and spatial length). The parameter $\delta$ represents a dimensionless wave amplitude, whereas the parameters $\text{g}^{(\alpha)}$, $\text{h}^{(\alpha)}$, $\varrho^{(\alpha)}$ are the dimensionless counterparts of the constants $g^{(\alpha)}$, $h^{(\alpha)}$, $\rho^{(\alpha)}$ introduced in Eqs.~\eqref{WaveReal}\textsubscript{1}-\eqref{ShearNL}. The latter have been normalised with respect to an effective elastic modulus $\langle g \rangle$ and an effective mass density $\langle \rho \rangle$, whose expressions are given in Table~\ref{tab:Effective}. In fact, these effective parameters govern linear wave propagation in the laminate \citep{andrianov13}, and the squared characteristic wave speed satisfies $\mathfrak{c}^2 = \langle g \rangle / \langle \rho \rangle$.

Then, we introduce a small parameter $\epsilon = \ell / \mathfrak{L}$ which expresses the smallness of the spatial period $\ell$ with respect to the macroscopic wavelength, $\mathfrak{L}$, see Fig.~\ref{fig:Struct}b for the notations. The parameter $\epsilon$ is then used for the asymptotic analysis of the composite in the long wave limit, $\mathfrak{L} \gg \ell$.
We describe the spatial evolution of the unknown displacement $\text{u}$ in terms of a `slow' spatial coordinate $\text{y}$ and of a new `fast' spatial coordinate, $\tilde{\text y} = \text{y}/\epsilon$. This way, the spatial derivative $\partial_\text{y} = (\cdot)_\text{y}$ with respect to $\text y$ must be replaced by the operator $(\cdot)_\text{y} + \frac{1}{\epsilon} (\cdot)_{\tilde{\text y}}$ in the wave equation \eqref{WaveFinal}\textsubscript{1}, which becomes
\begin{equation}
	\left( \text{g}^{(\alpha)} + \text{h}^{(\alpha)} \delta^2 (\text{u}_\text{y} + \tfrac{1}{\epsilon} \text{u}_{\tilde{\text y}} )^2 \right) (\partial_\text{y} + \tfrac{1}{\epsilon} \partial_{\tilde{\text y}})^2 \text{u} = \varrho^{(\alpha)} \text{u}_\text{tt} .
	\label{WaveEps}
\end{equation}
The unknown displacement $\text{u}(\text{y},\tilde{\text y},\text{t})$ is sought in the form of a power series of the small parameter $\epsilon$, i.e.,
\begin{equation}
	\text{u} = \text{u}_{0} + \epsilon \text{u}_{1} + \epsilon^2 \text{u}_{2} + \epsilon^3 \text{u}_{3} + \dots ,
	\label{Power}
\end{equation}
which is 1-periodic with respect to the fast coordinate $\tilde{\text y}$. The leading order homogenised term $\text{u}_0(\text{y},\text{t})$ depends only on the time coordinate and on the slow spatial coordinate, whereas higher-order corrections $\epsilon^n \text{u}_n(\text{y},\tilde{\text y},\text{t})$ depend on the fast spatial coordinate as well.

For convenience, we set the origin of the fast coordinates $\tilde{\text y}$ in the center of the periodic unit cell (dotted line in Fig.~\ref{fig:Struct}b). At a material interface between two bonded layers with distinct material properties (that is, at $\tilde{\text y} = \pm \frac12 \nu^{(2)}$ and at $\tilde{\text y} = \pm \frac12$), the displacement and the normal stress are required to be continuous. These conditions lead to the continuity of both expressions
\begin{equation}
	\text{u},
	\quad
	\text{g}^{(\alpha)} (\text{u}_\text{y} + \tfrac1{\epsilon} \text{u}_{\tilde{\text y}}) + \tfrac13 \text{h}^{(\alpha)} \delta^2 (\text{u}_\text{y} + \tfrac1{\epsilon} \text{u}_{\tilde{\text y}})^3,
	\label{StressEps}
\end{equation}
across such an interface. Using the power series assumption \eqref{Power} and splitting the boundary-value problems resulting from \eqref{WaveEps}-\eqref{StressEps} with respect to powers of $\epsilon$, we obtain a sequence of boundary-value problems for the displacement fields $\text{u}_n$ that can be restricted to a unit cell, $-\frac12 \leq \tilde{\text y} \leq \frac12$, by virtue of periodicity. A solution for $\text{u}_1$ is presented in the \ref{app:Homogen}, where the function $\text{u}_1$ is normalised in such a way that its integral with respect to the fast coordinate $\tilde{\text y}$ equals zero. In a similar fashion to \citet{andrianov13}, the fields $\text{u}_2$, $\text{u}_3$ are sought in the linear approximation, for which detailed expressions are provided by \citet{andrianov08} or by \citet{cornaggia20}.

Upon inserting the power series solution \eqref{Power} so-obtained into the wave equation \eqref{WaveEps} and applying the homogenising operator $\int_{-1/2}^{1/2} (\cdot) \, \text{d}\tilde{\text y}$ to the remaining leading-order term, we arrive at the macroscopic wave equation of order $(\epsilon^2, \delta^2)$, which takes the form
\begin{equation}
	\left(1 + \zeta \delta^2 (\partial_\text{y} \text{u}_0)^2\right) \partial_\text{yy} \text{u}_0 + \eta \epsilon^2 \partial_\text{y}^4 \text{u}_0 = \partial_\text{tt} \text{u}_0 ,
	\label{WaveHom}
\end{equation}
where the displacement field $\text{u}_0$ can be replaced with $\text u$ at the same order of approximation, see Eq.~\eqref{Power}.
This wave equation is valid for a moderate scale separation parameter $\epsilon$ and a moderate wave amplitude $\delta$. 
The effective coefficients are given by the formulas
\begin{equation}
	\begin{aligned}
		\zeta &= \frac{\nu^{(1)} \text{h}^{(1)} (\text{g}^{(2)})^4 + \nu^{(2)} \text{h}^{(2)} (\text{g}^{(1)})^4}{(\nu^{(1)} \text{g}^{(2)} + \nu^{(2)} \text{g}^{(1)})^4} , \\
		\eta &= \frac{1}{12} \frac{(\nu^{(1)} \nu^{(2)})^2}{(\text{g}^{(1)} \text{g}^{(2)})^2} \left( \varrho^{(1)}\text{g}^{(1)} -\varrho^{(2)}\text{g}^{(2)} \right)^2 .
	\end{aligned}
\end{equation}
Here, we have used the expressions in Eq.~\eqref{WaveFinal} and Table~\ref{tab:Effective} to convince ourselves that the macroscopic counterparts of the coefficients $\text{g}^{(\alpha)}$ and $\varrho^{(\alpha)}$ are equal to unity. Transforming back to dimensional variables and coordinates using Eq.~\eqref{WaveFinal}, we obtain the final form of the macroscopic wave equation \eqref{WaveEff}.

The wave equation \eqref{WaveEff} includes the nonlinear effect of moderate strain amplitudes (for $\zeta \neq 0$) and the effect of heterogeneity on wave dispersion characteristics (for $\eta \neq 0$). Through partial differentiation with respect to $y$, a direct consequence of our result is that the shear strain is governed by a Boussinesq-type wave equation \citep{christov07}.

\subsection{Towards a homogenised material theory}

Let us focus on the nonlinear elasticity part. On the one hand, according to our earlier analysis (Section~\ref{sec:Laminate}), the quasi-static response of the laminate in tension-compression is equivalent to that of a homogeneous material described by the average energy function, $\overline{W} = \nu^{(1)}W^{(1)} + \nu^{(2)}W^{(2)}$. On the other hand, the asymptotic homogenisation procedure of the present section provides an effective nonlinear shear wave equation for the periodic laminate, which accounts for the effect of tensile or compressive stresses in the lamination direction. It remains to determine whether it is possible to reconcile these two results, in order to obtain a unified homogenisation result.

This process is well-established in the case of neo-Hookean laminates, i.e., for layers described by a strain energy function of the form $W^{(\alpha)} = \frac12 \mathcal{G}^{(\alpha)} (I_1-3)$. In this case, a suitable expression of the effective strain energy reads \citep{galich17}
\begin{equation}
	W^\text{eff} = \tfrac12 \overline{\mathcal G} (I_1-3) - \tfrac12 (\overline{\mathcal G} - \breve{\mathcal G}) K ,
	\label{Galich}
\end{equation}
where $K = |\bm{F}\bm{n}|^2 - |\bm{F}^{-\text{T}}\bm{n}|^{-2}$. Here, the unit vector $\bm n$ indicates the initial lamination direction, and the coefficients
\begin{equation}
	\overline{\mathcal G} = \nu^{(1)}\mathcal{G}^{(1)} + \nu^{(2)}\mathcal{G}^{(2)}, \quad
	\breve{\mathcal G} = \left(\frac{\nu^{(1)}}{\mathcal{G}^{(1)}} + \frac{\nu^{(2)}}{\mathcal{G}^{(2)}}\right)^{-1}
	\label{GalichCoeff}
\end{equation}
are the arithmetic and the harmonic average shear moduli. For this strain energy function, the second term vanishes under tension-compression along the lamination direction, so that $W^\text{eff} = \frac12 \overline{\mathcal G} (I_1-3)$ is equal to the linear average $\overline{W}$ of the phase-related strain energy contributions. In addition, the shearing motion \eqref{FShear} along the layer interfaces is governed by the partial differential equation
\begin{equation}
	\lambda^2 \breve{\mathcal G} u_{yy} = \langle g\rangle u_{yy} = \langle\rho\rangle u_{tt} ,
	\label{GalichWave}
\end{equation}
which is coherent with our homogenisation result \eqref{WaveEff}, in the limit of long waves with infinitesimal amplitude. To arrive at this conclusion, we have used the definition of the coefficients $g^{(\alpha)}$ of Table~\ref{tab:Elast} where we have set $\beta^{(\alpha)} = 0$, while the effective mass density was taken equal to $\langle\rho\rangle$.

The definition of a suitable effective strain energy for highly nonlinear laminates with Yeoh, Fung-Demiray, or Gent phases is more challenging, and analytical expressions are not easily obtained. In the case of hyperelastic phases described by the Gent theory, \citet{yao24} recently followed a heuristic approach.\footnote{There is possibly an error in the expression of the shear moduli (24) and (27) in \citet{yao24}, see their definition in Eq.~\eqref{GalichCoeff}.} Their suggestion is to substitute the shear moduli $\mathcal{G}^{(\alpha)}$ in Eqs.~\eqref{Galich}-\eqref{GalichCoeff} by the generalised Gent shear moduli ${G}^{(\alpha)}$ (see expression in Table~\ref{tab:Elast}), where the the first invariant $I_1$ is taken equal to its average value $I_1^{(\alpha)}$ in a given phase. Nevertheless, such an \emph{ad hoc} theory seems hardly applicable in the present case. 

In the \ref{app:LopezPamies}, we demonstrate that for small values of the parameters $\beta^{(\alpha)}$, a suitable effective strain energy for Yeoh, Fung-Demiray, or Gent laminates reads
\begin{equation}
	\begin{aligned}
	W^\text{eff} &= \tfrac12 \overline{\mathcal G} D + \tfrac12 \breve{\mathcal G} K \\
	&\quad + \tfrac14 \overline{\mathcal{G}^1\beta} D^2 + \tfrac12 \breve{\mathcal G}^2 \overline{\mathcal{G}^{-1}\beta} D K + \tfrac14 \breve{\mathcal G}^4\overline{\mathcal{G}^{-3}\beta} K^2 ,
	\end{aligned}
	\label{Weff}
\end{equation}
where $D = I_1 - 3 -K$, and $K$ was introduced in Eq.~\eqref{Galich}. Here, we have introduced the averages
\begin{equation}
	\overline{\mathcal{G}^n\beta} = \nu^{(1)} (\mathcal{G}^{(1)})^n \beta^{(1)} + \nu^{(2)} (\mathcal{G}^{(2)})^n \beta^{(2)} ,
	\label{AvNote}
\end{equation}
defined for any integer $n$. In particular, we note that this expression is consistent with Eq.~\eqref{Galich} in the neo-Hookean limit, for which $\beta^{(\alpha)} \to 0$ and all the coefficients $\overline{\mathcal{G}^n\beta}$ vanish. For a homogeneous material ($\nu^{(1)} = 1$), the strain energy function \eqref{Weff} reduces to that of the isotropic Yeoh theory presented in Table~\ref{tab:Elast} with $\alpha=1$. Nevertheless, an effective theory for highly nonlinear soft laminates with arbitrary values of the coefficients $\beta^{(\alpha)}$ remains to be found.

The procedure described in the \ref{app:LopezPamies} could be used to derive relevant effective magneto-active material theories, see for instance the expression in \citet[Eqs.~(27)-(28)]{karami19}, as well as the derivations in \citet{spinelli15}. Along the same lines, the derivation of a suitable three-dimensional theory that accounts for wave dispersion is an open problem, to the authors' present knowledge. Potential anisotropic macroscopic theories that incorporate these microstructural effects might belong to a specific class of dispersive solids \citep{destrade08}, or of generalised continuum models, such as the micromorphic, Cosserat, or the second-gradient elasticity theory \citep{madeo15, forest20}.

\section{Nonlinear and dispersive waves}\label{sec:NLWaves}

In this section, we will investigate various properties of the propagation of shear waves in soft laminates. Our starting point is the equation of motion \eqref{WaveEff}, which is deduced from the asymptotic homogenisation procedure. We derive a nonlinear dispersive wave equation which shares many properties with a related equation that was used to study wave propagation in polystyrene bars \citep{garbuzov22}.

\subsection{Small amplitude waves}

We first focus on waves of small amplitude, for which the nonlinear term can be neglected in the equation of motion. Thus, we set $\zeta = 0$ in Eq.~\eqref{WaveEff}, or equivalently, the layers are assumed neo-Hookean ($\beta^{(\alpha)} = 0$). Regarding our homogenisation procedure, it is worth mentioning that the description of the low-frequency dispersion characteristics can be improved by replacing the fourth-order spatial derivative in Eq.~\eqref{WaveEff} as follows,
\begin{equation}
	\begin{aligned}
	&\eta u_{yyyy} \simeq \eta_y u_{yyyy} - \frac{\eta_m}{\mathfrak{c}^2} u_{yytt} - \frac{\eta_t}{\mathfrak{c}^4} u_{tttt} , \\
	&\eta = \eta_y - \eta_m - \eta_t ,
	\end{aligned}
	\label{Wautier}
\end{equation}
where $\mathfrak{c}^2 = \langle g\rangle / \langle \rho\rangle$, which is equivalent to the initial dispersive term $\eta u_{yyyy}$ at the same order of approximation \citep{wautier15}. As shown hereinafter, the coefficients $\eta_y$, $\eta_m$, $\eta_t$ can then be chosen in such a way that the modified equation of motion provides an accurate approximation of the exact dispersion relationship of the layered material. Within this framework, the effective equation of motion \eqref{WaveEff} is recovered by setting $\eta_m = \eta_t = 0$.

To enable such comparisons, let us recall that the propagation of a harmonic wave of the form $u \propto \text{e}^{\text{i} (\kappa y - \omega t)}$ in the layered material is governed by the exact dispersion relationship
\begin{equation}
	\begin{aligned}
	&\cos (\kappa \ell) =  \cos \left(\frac{\omega \ell^{(1)}}{\mathfrak{c}^{(1)}}\right) \cos \left(\frac{\omega \ell^{(2)}}{\mathfrak{c}^{(2)}}\right) \\
	&\; - \frac12 \left(\frac{\rho^{(1)} \mathfrak{c}^{(1)}}{\rho^{(2)} \mathfrak{c}^{(2)}} + \frac{\rho^{(2)} \mathfrak{c}^{(2)}}{\rho^{(1)} \mathfrak{c}^{(1)}} \right) \sin \left(\frac{\omega \ell^{(1)}}{\mathfrak{c}^{(1)}}\right) \sin \left(\frac{\omega \ell^{(2)}}{\mathfrak{c}^{(2)}}\right)  ,
	\end{aligned}
	\label{DispExact}
\end{equation}
where $\kappa$ is the wave number, $\omega$ is the angular frequency, $\mathfrak{c}^{(\alpha)} = \sqrt{g^{(\alpha)} / \rho^{(\alpha)}}$ is the wave speed in the phase $\alpha$, and $\ell^{(\alpha)} = \nu^{(\alpha)} \ell$ is the thickness of its layers. This dispersion relationship is obtained by following the Floquet--Bloch approach, see \cite{andrianov08} for details. The propagation of such a perturbation in the homogenised medium \eqref{WaveEff}-\eqref{Wautier} satisfies
\begin{equation}
	\eta_t \left(\frac{\omega \ell}{\mathfrak{c}}\right)^4 - \left(1 - \eta_m (\kappa \ell)^2\right) \left(\frac{\omega \ell}{\mathfrak{c}}\right)^2 + (\kappa \ell)^2 - \eta_y (\kappa \ell)^4 = 0 .
	\label{DispHom}
\end{equation}
As a special case, we find that $({\omega \ell}/{\mathfrak{c}})^2 = (\kappa \ell)^2 - \eta (\kappa \ell)^4$ for the effective equation of motion \eqref{WaveEff}, which can be viewed as a low-order approximation of the exact dispersion relationship \eqref{DispExact}, see for instance \citet{cornaggia20}. To obtain a more accurate representation of wave dispersion,
it remains to adjust the coefficients $\eta_y$, $\eta_m$, $\eta_t$ in a suitable manner, for which various strategies can be followed. However, some properties of the exact dispersion relationship \eqref{DispExact} will not be recovered by using the homogenised theory \eqref{DispHom}, such as the $2\pi$-periodicity of the frequency with respect to $\kappa \ell$, or the reflection symmetry of the frequency at $\kappa \ell = 0$ and at $\kappa \ell = \pi$.

\begin{figure}
	\centering
	
	\includegraphics{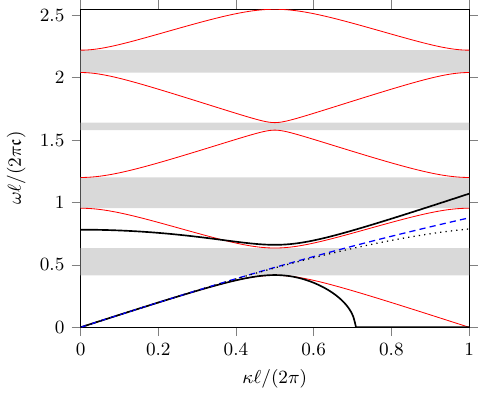}
	
	\caption{Dispersion curves deduced from the exact dispersion relationship \eqref{DispExact} (thin lines), from the homogenised model \eqref{WaveEff} (dotted line), and from the homogenised model \eqref{DispHom} with the optimised coefficients \eqref{WautierParam} (thick black lines). The dashed curve marks the dispersion relationship \eqref{DispKdV} of unidirectional waves. Here, the laminate is made of soft neo-Hookean phases with a high shear modulus contrast ($\mathcal{G}^{(1)}=5\mathcal{G}^{(2)}$), and the material is in its undeformed configuration. \label{fig:Disp}}
\end{figure}

Among such models, the `good' Boussinesq equation where $\eta_y$ and $\eta_t$ equal zero was used by \citet{fish02}. This equation can be obtained directly based on a modified homogenisation procedure \citep{vivar09}. Alternatively, as suggested by \citet{cornaggia23}, we might restrict our attention to the class of dispersive models for which $\eta_y=0$, whose two parameters can be adjusted to match the exact dispersive behaviour of the laminate with a satisfactory accuracy in the long-wave limit. In addition, an optimised set of parameters corresponding to the case $\eta_t=0$ is provided by \citet{wautier15}, as well as an optimisation technique for the full three-parameter model \eqref{Wautier}. According to the latter, the three parameters can be adjusted to match closely the first frequency band gap of the periodic structure (Figure~\ref{fig:Disp}).

To do so, we adapt the optimisation technique of \citet{wautier15} as follows:
\begin{itemize}
\item[(i)] First, we note that a positive value of $\eta_t$ in Eq.~\eqref{DispHom} produces two distinct real-valued solutions for $(\omega \ell/\mathfrak{c})^2$ in the long-wave limit $(\kappa\ell \to 0)$, whereas a single real solution is obtained for $\eta_t \leq 0$. We will therefore focus on the case $\eta_t > 0$, and rewrite the dispersion relationship \eqref{DispHom} divided by $\eta_t$ in the form
\begin{equation}
	 \chi^2 + 2\phi\chi + \psi = 0, \quad \chi = ({\omega \ell}/{\mathfrak{c}})^2 ,
	 \label{WautierQuad}
\end{equation}
with suitable coefficients $\phi$ and $\psi$. Note that these coefficients and variable should not be confused with the energy functions introduced in \eqref{Steigmann}-\eqref{Legendre}, as well as with the magnetic susceptibility introduced in \eqref{MagRelations}.
\item[(ii)] Then, the group velocity $\frac{\text d \omega}{\text d \kappa}$ is required to equal zero for $\kappa \ell = \pi$ on all branches of the solution. An equivalent requirement is that $\frac{\text d \chi}{\text d \kappa}|_{\kappa \ell = \pi}$ vanishes, if we assume that the corresponding frequency is non-zero. Differentiating the solutions of \eqref{WautierQuad} with respect to the wave number, we deduce that both $\frac{\text d \phi}{\text d \kappa \ell}$ and $\frac{\text d \psi}{\text d \kappa \ell}$ must equal zero for $\kappa \ell = \pi$, which implies that
\begin{equation}
	\eta_y = \frac{1}{2 \pi^2} , \quad \eta_m = 0, \quad \eta_t = \frac{1}{2 \pi^2} - \eta > 0 .
	\label{WautierParam}
\end{equation}
The parameter $\eta_t$ is deduced from \eqref{Wautier}\textsubscript{2} with known values of $\eta_y$, $\eta_m$ and $\eta$ (Table~\ref{tab:Effective}).
\end{itemize}
Incidentally, this model is actually a two-parameter model for which $\eta_m = 0$. Given the approximation $\eta_y \approx 0.0507$, we were able to reach the same conclusion as \citet{wautier15} without making use of numerical optimisation (see Fig.~6 therein).

To illustrate, let us reproduce the computations leading to Fig.~5a of \citet{zhang22}. Thus, we consider wave numbers within the interval $0 \leq \kappa\ell\leq 2\pi$, and we select the material parameters $\nu^{(1)} = \frac12$, $\rho^{(1)} = \rho^{(2)}$, and $\mathcal{G}^{(2)}/\mathcal{G}^{(1)} = \frac15$. This shear modulus ratio differs from that chosen by \citet{zhang22}, $\mathcal{G}^{(2)}/\mathcal{G}^{(1)} = 15$, for the sake of material feasibility. The material is in its initial and undeformed state, so that $\lambda = 1$. On the one hand, we calculate the angular frequency $\omega$ by solving the exact dispersion relationship \eqref{DispExact} numerically, while on the other hand, we calculate the angular frequency $\omega$ deduced from the homogenised dispersion equation \eqref{DispHom} with the parameters \eqref{WautierParam}. Here, we find $\eta \approx 0.00926$, so that $\eta_y \approx 0.0507$ and $\eta_t \approx 0.0414$.

The results are reported in Figure~\ref{fig:Disp} where the dispersion curves are compared. Based on the exact dispersion curves, we observe the emergence of frequency band gaps where waves cannot propagate due to destructive interference (shaded areas). This feature is not reproduced by the effective wave equation \eqref{WaveEff} where $\eta_m = \eta_t = 0$. Nevertheless, the first band gap is reasonably well captured by the homogenised model with the optimised parameters $(\eta_y,\eta_m,\eta_t)$, which is consistent with the low-frequency assumption made in the previous section. In fact, inspection of the homogenised dispersion relationship \eqref{WautierQuad}-\eqref{WautierParam} shows that waves cannot propagate in the frequency interval
\begin{equation}
	0.84\pi \approx \sqrt{\frac{1 - \pi \sqrt{2\eta}}{2 \eta_t}} \leq \frac{\omega \ell}{\mathfrak{c}} \leq \sqrt{\frac{1 + \pi \sqrt{2\eta}}{2 \eta_t}} \approx 1.32\pi .
	\label{BandGap}
\end{equation}
Instead, the exact dispersion relationship \eqref{DispExact} yields the bounds $0.83\pi \leq \omega \ell/\mathfrak{c} \leq 1.27\pi$ for the first band gap. Thus, as displayed in the figure, the homogenised theory yields an accurate estimation of the first cut-off frequency, whereas the width of the first band gap is slightly overestimated. To improve the accuracy of homogenisation theories at higher frequency, dedicated procedures and effective theories have been developed, see for instance \citet{craster10} and \citet{willis09}.

\subsection{Unidirectional nonlinear waves}

We now move on to the study of nonlinear wave solutions, i.e., the laminate can no longer be described by purely neo-Hookean strain energy functions anymore. The propagation of shear waves is governed by the wave equation \eqref{WaveEff} with the dispersion terms \eqref{Wautier} which we recall here,
\begin{equation}
	\begin{aligned}
		\frac{u_{tt}}{\mathfrak{c}^2} &= \left(1 + \zeta u_y^2 \right) u_{yy}  \\
		&\quad + \ell^2 \left(\eta_y u_{yyyy} - \frac{\eta_m}{\mathfrak{c}^2} u_{yytt} - \frac{\eta_t}{\mathfrak{c}^4} u_{tttt}\right) ,
	\end{aligned}
	\label{WaveEffWautier}
\end{equation}
see Table~\ref{tab:Effective} and Eq.~\eqref{WautierParam} for the determination of the parameters.
In a similar fashion to \citet{zabolotskaya04}, we proceed to a reduction of the equation of motion for unidirectional waves with a slowly varying profile in space. This procedure is commonly applied to study waves that propagate in a preferred direction \citep{destrade08}. Moreover, it is consistent with the assumptions of our asymptotic homogenisation procedure, which is valid for moderate wave amplitudes and frequencies.

We introduce the \emph{scaling} defined by the change of variables $\lbrace \hat{y} = \varepsilon^2 y, \hat{t} = t-y/\mathfrak{c}, u = \varepsilon \hat{u}(\hat y, \hat t)\rbrace$, where $\mathfrak{c}$ is the wave speed and $\varepsilon$ is a small parameter of same order as the microstructure's characteristic length, $\ell$. This Ansatz is then substituted into the equation of motion \eqref{WaveEffWautier}. At leading (cubic) order in $\varepsilon$, the shearing motion is governed by a nonlinear scalar partial differential equation. Transforming back to the displacement $u(y,t)$ leads to a reduced wave equation for the shearing velocity $v = u_t$,
\begin{equation}
	v_y + \mathfrak{c}^{-1} \left(1 - \tfrac12 \zeta v^2/\mathfrak{c}^2\right) v_t - \tfrac12 \mathfrak{c}^{-3} \eta\ell^2 v_{ttt} = 0 , \quad
	\label{mKdV}
\end{equation}
which is a modified Korteweg--De Vries (mKdV) equation.

At this stage, it is worth noting that the same slow-space wave equation \eqref{mKdV} is obtained independently of the form of the fourth-order dispersion terms in the equation of motion, as long as the condition \eqref{Wautier}\textsubscript{2} on $(\eta_y,\eta_m,\eta_t)$ is fulfilled. In addition, the dispersion relationship deduced from Eq.~\eqref{mKdV} with $\zeta = 0$ reads
\begin{equation}
	\kappa\ell = \frac{\omega \ell}{\mathfrak{c}} + \frac{\eta}{2} \left(\frac{\omega \ell}{\mathfrak{c}}\right)^3 ,
	\label{DispKdV}
\end{equation}
which is represented by a dashed line in Fig.~\ref{fig:Disp}. This dispersion relationship is not equivalent to the dispersion relationship \eqref{DispHom} deduced from the equation of motion, for which wave dispersion is sensitive to the value of the coefficients $(\eta_y,\eta_m,\eta_t)$. This mismatch is a known feature of slow scale approximations \citep{berjamin23}.

\begin{remark}
A similar derivation can be performed by considering slowly varying wave profiles in the time domain \citep{berjamin23}. This way, we arrive at the following partial differential equation for the shear strain $\gamma = u_y$,
\begin{equation}
	 \gamma_t + \mathfrak{c} \left(1 + \tfrac12 \zeta \gamma^2\right) \gamma_y + \tfrac12 \mathfrak{c} \eta\ell^2 \gamma_{yyy} = 0 ,
	 \label{mKdV2}
\end{equation}
which is also of mKdV type, and has similar properties to Eq.~\eqref{mKdV}.
\end{remark}

\subsection{Solitary waves}

In a similar fashion to \citet{andrianov13}, let us now seek travelling wave solutions to the wave equation \eqref{WaveEffWautier}. Such solutions are smooth waveforms that propagate at a constant velocity with a steady profile. Thus, we assume that $u$ is a smooth function of the phase variable $\xi = (y-\mathfrak{s} t)/\ell$ only, where $\mathfrak{s}$ is the velocity of the wave. We introduce also the strain variable $\Phi$ such that $u' = \Phi \ell \sqrt{3/\zeta}$, where the prime denotes differentiation with respect to $\xi$. Insertion of these assumptions in the wave equation \eqref{WaveEffWautier} and integration with respect to $\xi$ produces a nonlinear oscillator equation,
\begin{equation}
	 \Phi'' + c_0 - c_1 \Phi + c_3 \Phi^3 = 0 ,
	 \label{SolitonODE}
\end{equation}
where $c_0$ is an arbitrary integration constant, and
\begin{equation}
	c_1 = (\mathfrak{s}^2/\mathfrak{c}^2-1) c_3, \quad c_3 = \frac{1}{\eta_y - \eta_m \mathfrak{s}^2/\mathfrak{c}^2 - \eta_t \mathfrak{s}^4/\mathfrak{c}^4} .
	\label{SolitonWav}
\end{equation}
Inserting the same Ansatz in Eq.~\eqref{mKdV}, we arrive at the same type of differential equation \eqref{SolitonODE}, with the coefficients
\begin{equation}
	c_1 = \frac{2 (\mathfrak{s}/\mathfrak{c}-1)}{\mathfrak{s}^3/\mathfrak{c}^3} c_3, \quad c_3 = \frac{1}{\eta} .
	\label{SolitonKdV}
\end{equation}
Among other solutions (see for instance \citet{el17} or \citet{ling22}), Eq.~\eqref{SolitonODE} admits a family of solitary waves of the form
\begin{equation}
	\Phi(\xi) = \pm\sqrt{\frac{2 c_1}{c_3}} \operatorname{sech} (\xi \sqrt{c_1}) ,
	\label{SolitonSol}
\end{equation}
where $\operatorname{sech} = 1/\cosh$ is the hyperbolic secant function. This solution is well-defined for $c_0 = 0$, $c_1 > 0$, and $c_3 > 0$, and the waveforms satisfy
\begin{equation}
	\begin{aligned}
	&u'/\ell = \delta \operatorname{sech} (\ell\xi / \mathfrak{L}), \quad 
	u = \mathfrak{a} \operatorname{gd}(\ell\xi / \mathfrak{L}) , \\
	&\mathfrak{L} = \frac{\ell}{\sqrt{c_1}}, \quad \mathfrak{a} = \ell \sqrt{\frac{6}{c_3 \zeta}}, \quad \delta = \sqrt{\frac{6 c_1}{c_3 \zeta}} ,
	\end{aligned}
	\label{Soliton}
\end{equation}
where the `+' sign in \eqref{SolitonSol} was selected, $\operatorname{gd}$ denotes the Gudermannian function defined by $\operatorname{gd}(\xi) = \arctan(\sinh \xi)$, and $u'/\ell = \gamma = -v/\mathfrak{s}$ is the shear strain. Here, $\mathfrak L$ denotes the characteristic length of the soliton \eqref{SolitonSol}, $\mathfrak a$ is the amplitude of the displacement field, and $\delta = \mathfrak{a}/\mathfrak{L}$ is the characteristic strain amplitude. It can be easily verified that the above functions \eqref{Soliton} are special solutions of the wave equations \eqref{WaveEffWautier} and \eqref{mKdV}-\eqref{mKdV2} with suitable values of the coefficients $c_1$ and $c_3$.

From the expression of the wave amplitude, we can infer the relationship between the amplitude of a solitary wave and its velocity. In particular, we note that these relationships are not identical depending on whether the whole equation of motion \eqref{WaveEffWautier} or its slow space approximation \eqref{mKdV} was used. Nevertheless, in both cases, we find that $\frac16 \delta^2 \zeta = c_1/c_3$ is asymptotically equivalent to $2\, (\mathfrak{s}/\mathfrak{c}-1)$ in the vicinity of $\mathfrak{s}=\mathfrak{c}$. In other words, for a given value of $\zeta$, the soliton's velocity increases proportionally to the square of its strain amplitude, within a range of moderate wave amplitudes.

A major difference between the wave equation \eqref{WaveEffWautier} and the slow-scale approximation \eqref{mKdV} lies in the sign of the coefficients $c_3$ defined above. In fact, the coefficient $c_3$ deduced from the slow-scale approximation is positive, so that solitons with relative velocity $\mathfrak{s}>\mathfrak{c}$ always exists. However, the picture is very much different in the case of the wave equation \eqref{WaveEffWautier}, which admits solitary wave solutions provided that the denominator of $c_3$ remains positive for $\mathfrak{s}>\mathfrak{c}$, or equivalently that
\begin{equation}
	\eta - (\eta_m+2\eta_t)\vartheta^2 - \eta_t \vartheta^4 > 0 , \quad \vartheta^2 = \mathfrak{s}^2/\mathfrak{c}^2-1 > 0 ,
	\label{IneqSol}
\end{equation}
where we have used \eqref{Wautier}\textsubscript{2}.
This requirement is satisfied for all $\mathfrak{s}>\mathfrak{c}$ if either:
\begin{itemize}
	\item[(i)] the polynomial function \eqref{IneqSol} of the variable $\vartheta^2$ is degenerate ($\eta_t = 0$), and $\eta_m \leq 0$. From \eqref{Wautier}\textsubscript{2}, it follows that $\eta_y \leq \eta$.
	\item[(ii)] the polynomial function \eqref{IneqSol} is not degenerate ($\eta_t \neq 0$), and it has only real roots which are negative. Therefore, the discriminant satisfies $(\eta_m+2\eta_t)^2 + 4 \eta_t \eta \geq 0$. Furthermore, the sum of the roots must be non-positive, $-(\eta_m+2\eta_t)/\eta_t \leq 0$, and their product must be non-negative, $-\eta/\eta_t \geq 0$. From these conditions, we conclude that $\eta_t < 0$, $\eta_m \leq -2\eta_t$, and $\eta_y \leq -\frac14\eta_m^2/\eta_t$, where we have used \eqref{Wautier}\textsubscript{2}.
	\item[(iii)] the polynomial function \eqref{IneqSol} is not degenerate ($\eta_t \neq 0$), and it has no real root, i.e., $(\eta_m+2\eta_t)^2 + 4 \eta_t \eta < 0$. In particular, this implies that $\eta_t < 0$. Using \eqref{Wautier}\textsubscript{2}, the above requirement may also be rewritten as $\eta_y > -\frac14\eta_m^2/\eta_t$.
\end{itemize}
Otherwise, there exists a range of relative wave speeds $\mathfrak{s}/\mathfrak{c}$ for which the solitary wave \eqref{SolitonSol} cannot propagate, and this range corresponds to specific wavelengths and amplitudes.

In particular, the inequality \eqref{IneqSol} is naturally satisfied in the case where both $\eta_m$ and $\eta_t$ are equal to zero. However, as shown above, in the special case $\eta_y=0$ with $\eta_t$ positive, the above inequality yields an upper bound for the relative wave velocity, $\mathfrak{s}^2/\mathfrak{c}^2-1 < \eta/\eta_t$, where use was made of \eqref{Wautier}\textsubscript{2}. Similarly, in the special case $\eta_t=0$, the above inequality yields the upper bound $\mathfrak{s}^2/\mathfrak{c}^2-1 < \eta/\eta_m$, unless the coefficient $\eta_m$ is non-positive. In the special case $\eta_m = 0$ with $\eta_t$ positive (see Eq.~\eqref{WautierParam}), the inequality \eqref{IneqSol} yields the upper bound $\mathfrak{s}^2/\mathfrak{c}^2 < \sqrt{1 + \eta/\eta_t}$.

In addition, there are some further differences between the properties of the $\operatorname{sech}$-solitons deduced from the wave equation \eqref{WaveEffWautier} and those deduced from the mKdV equation \eqref{mKdV}. In fact, for the former, the squared strain amplitude $\delta^2$ is an even function of the velocity $\mathfrak{s}$ which increases with $|\mathfrak{s}|$, whereas these properties do not hold for the latter. It turns out that the squared amplitude $\delta^2$ is not an even function of the relative velocity in this case{\,---\,}this is coherent with the fact that the mKdV equation describes \emph{unidirectional} waves propagating towards increasing $y${\,---\,}and that $\delta^2$ is only increasing up to $\mathfrak{s}/\mathfrak{c} = 1.5$ where it reaches a local maximum. For a more quantitative comparison, we note that the relative error in the strain amplitude $\delta$ introduced by the slow space approximation remains below $10\%$ for $\mathfrak{s} < 1.06\, \mathfrak{c}$ only. These bounds give us a limit for the validity of the one-way model equation \eqref{mKdV}.

\begin{remark}
The same travelling wave analysis applied to the slow time approximate model \eqref{mKdV2} yields the differential equation \eqref{SolitonODE} with the coefficients $c_1 = 2 (\mathfrak{s}/\mathfrak{c}-1) c_3$ and $c_3 = 1/\eta$. In this case, the relationship between the wave speed and the wave amplitude is again equivalent to that deduced from the wave equation \eqref{WaveEffWautier} in the vicinity of $\mathfrak{s}=\mathfrak{c}$. However, the wave amplitude is an uneven function of the wave speed, and the relative error in the strain amplitude $\delta$ introduced by the slow time approximation does not exceed $10\%$ for $\mathfrak{s} < 1.46\, \mathfrak{c}$ only.
\end{remark}

\begin{table*}
	\caption{Physical parameters of an example bi-layered material with equal volume fractions $\nu^{(\alpha)} = \frac12$ in both phases, together with the effective parameters deduced from homogenisation theory when the material is undeformed (Tables~\ref{tab:Effective}-\ref{tab:Elast} where $\lambda = 1$ and $\mathcal{I}_1 = 3$). The linear dispersion parameters are deduced from Eq.~\eqref{WautierParam} with $\eta = 0.00926$. \label{tab:Param}}
	\vspace{0.5em}
	\centering
	{\renewcommand{\arraystretch}{1}
	\small
	\begin{tabular}{ccccccccc}
    \toprule 
    $\alpha$ & $\ell^{(\alpha)}$ & $\rho^{(\alpha)}$ & $\mathcal{G}^{(\alpha)}$ & $\mathfrak{c}^{(\alpha)}$ & $\beta^{(\alpha)}$ & & & \\
    \midrule
    $1$  & $0.5$~cm & $930$~kg/m\textsuperscript{3} & $4.7$~MPa & $71.1$~m/s & $0.0132$ & & & \\ 
    $2$  & $0.5$~cm & $930$~kg/m\textsuperscript{3} & $0.94$~MPa & $31.8$~m/s & $0.0132$ & & & \\ 
    \midrule
    Eff. & $\ell$ & $\langle \rho \rangle$ & $\langle g \rangle$ & $\mathfrak{c}$ & $\zeta$ & $\eta_y$ & $\eta_m$ & $\eta_t$ \\
    \midrule
     & $1.0$~cm & $930$~kg/m\textsuperscript{3} & $1.57$~MPa & $41.0$~m/s & $0.0924$ & $0.0507$ & $0$ & $0.0414$ \\
    \bottomrule 
    \end{tabular}}
\end{table*}

\begin{figure*}
	\centering
	\begin{minipage}{0.49\textwidth}
		\centering
		(a)
		
		\includegraphics{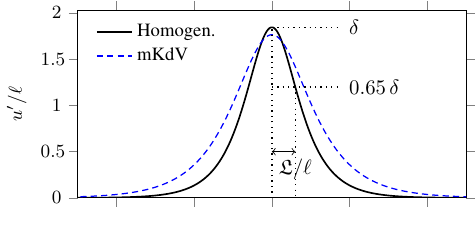}
		
		\vspace{-1em}
		
		\includegraphics{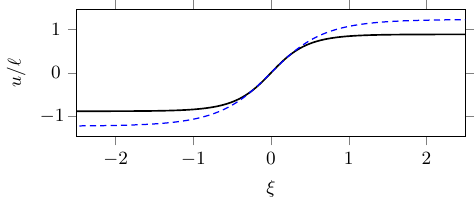}
	\end{minipage}
	\hfill
	\begin{minipage}{0.49\textwidth}
		\centering
		(b)
		
		\includegraphics{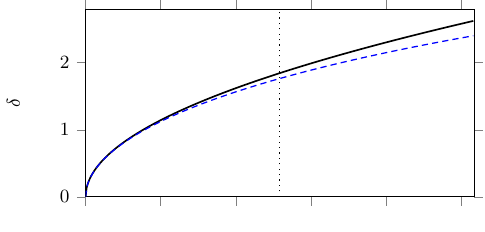}
		
		\vspace{-1em}
		
		\hspace{-0.6em}
		\includegraphics{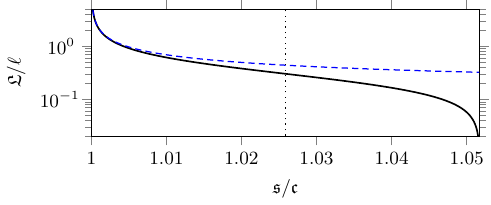}
	\end{minipage}

	\caption{Solitons. (a) Waveforms \eqref{Soliton} obtained with the parameters of Table~\ref{tab:Param} for $\mathfrak{s} = 1.026\, \mathfrak{c}$. The full homogenised theory (solid lines) is compared to the one-way mKdV-type model (dashed lines). For the homogenised theory, dotted lines highlight the strain amplitude $\delta$ and the characteristic wave length $\mathfrak{L}/\ell$, which can be deduced from the width of the wave at about $65\%$ of its amplitude. (b) Evolution of the wave amplitude (top) and of the wave length (bottom) in terms of the wave speed for both models. The vertical dotted lines mark the wave speed used in (a). \label{fig:Soliton}}
\end{figure*}

We illustrate these properties in Figure~\ref{fig:Soliton}a where the waveforms \eqref{Soliton} are represented based on suitable values of the coefficients $c_1$ and $c_3$, see Eqs.~\eqref{SolitonWav}-\eqref{SolitonKdV}. Although the solitary wave deduced from the mKdV equation \eqref{mKdV} appears similar to that obtained with the homogenised theory \eqref{WaveEffWautier}, the figure highlights an amplitude and wavelength mismatch between the two solutions. Here, the linear elastic properties of the stratified material are the same as for the study of linear waves (Fig.~\ref{fig:Disp}), i.e., we use the same volume fractions $\nu^{(\alpha)} = \ell^{(\alpha)}/\ell$ as well as the same ratios $\rho^{(2)}/\rho^{(1)}$ and $\mathcal{G}^{(2)}/\mathcal{G}^{(1)}$. We specify the reference parameters $\ell^{(1)}$, $\rho^{(1)}$ and $\mathcal{G}^{(1)}$ for one phase as described in Table~\ref{tab:Param}. We choose identical values of $\beta^{(\alpha)} = 1/76$ in both phases. For the phase $\alpha=1$, the parameter values in Table~\ref{tab:Param} are coherent with measurements obtained on rubber samples \citep[Table~5]{destrade17}, in which we have neglected the contribution of the second invariant, `$I_2$'.

Based on the effective parameters in Table~\ref{tab:Param} and the above analysis, we find that the upper bound $\sqrt{1 + \eta/\eta_t}$ for the ratio $\mathfrak{s}^2/\mathfrak{c}^2$ is equal to $1.11$ in the present case, when the homogenised theory \eqref{WaveEffWautier} is used. This bound corresponds to the maximum value of the wave speed on the horizontal axis in Fig.~\ref{fig:Soliton}b, where $\max \mathfrak{s}/\mathfrak{c} \approx 1.052$. We observe that the solitary wave solution \eqref{SolitonWav}-\eqref{Soliton} becomes singular as $\mathfrak{s}/\mathfrak{c}$ approaches this critical value, for which strain amplitudes remain finite, but the wavelength becomes infinitely small. Here, the strain amplitude cannot exceed
\begin{equation}
	\max\delta = \sqrt{\frac{\sqrt{1 + \eta/\eta_t}-1}{\zeta/6}} \approx 2.6.
	\label{SolitonDeltMax}
\end{equation}
Interestingly, there is no such singularity when the mKdV equation \eqref{mKdV} is used. In addition, it appears that both theories admit identical solitary wave solutions in the limit of small amplitudes and long wavelengths ($\mathfrak{s}/\mathfrak{c} \to 1$), as illustrated in Fig.~\ref{fig:Soliton}b.

\subsection{Impact problem}\label{subsec:Impact}

To validate our continuum theories, we aim at comparing them to direct numerical simulations. One problem of interest is the shear impact problem, in which a given shearing velocity is applied to the free boundary of a half-space. In the case of a homogeneous soft elastic solid, shock waves will develop at a finite distance from the boundary due to nonlinear distorsion (i.e., the particle velocity becomes discontinuous), and the shock formation distance can be estimated analytically \citep{chocka20}. This method was applied by \citet{conroy24} who studied shock formation in elastic laminates. Using the strain energy function \eqref{Galich} and neglecting wave dispersion caused by reflection on the layer interfaces at non-zero frequencies (i.e., implicitly assuming matching acoustic impedances), they found that the shock formation characteristics are highly dependent on the lamination direction, $\bm n$.

However, wave dispersion has a tremendous influence on the solutions of the impact problem. In fact, a study of nonlinear laminates by \citet{yong03} shows that soliton-like wave solutions might form instead of shocks. Related results were obtained in a variety of nonlinear layered media, with similar conclusions \citep{berjamin19, ziv20}. The emergence of these solitary waves (aka. dispersive shock waves) was studied by \citet{ketcheson12} using computer simulations and notions of entropy evolution, whereas \citet{purohit22} used modulation theory. Both studies conclude that stratified media tend to facilitate the formation of smooth waves, while inhibiting the formation of discontinuities. Nevertheless, shock waves can still be potentially generated at sufficiently large amplitudes.

Here, we consider a smooth impact problem defined by
\begin{equation}
	u_t(0, t) = \begin{cases}
		\mathscr{V} \sin^2(\frac12 \kappa \mathfrak{c} t) , \quad 0 \leq \kappa \mathfrak{c} t\leq 2\pi ,\\
		0 \quad \text{otherwise} .
	\end{cases}
	\label{Impact}
\end{equation}
The sample, which occupies the region $y\geq 0$, is initially undeformed and at rest. It is a layered medium governed by Eq.~\eqref{WaveReal}\textsubscript{1} in each phase whose parameters are specified in Table~\ref{tab:Param} (unless specified otherwise hereinafter), and the origin $y=0$ is located in the middle of a layer of material with phase index $\alpha=2$. To ensure separation of scales, we set the effective wavelength equal to sixteen spatial periods, i.e., the wave number satisfies $2\pi/\kappa = 16\, \ell$. Furthermore, we place ourselves in a range of wave amplitudes such that $\mathscr{V} = 2\mathfrak{c}$.

For the heterogeneous material, we implement a second-order finite volume scheme with limiters \citep{yong03,bale03}, see also \citet{fogarty99} where the definition of the limiter is provided. The numerical solution is computed using 32 finite volumes in each phase, so that in effect, our solution has 1024 points per wavelength. The time step is chosen in such a way that the Courant number equals 0.95 at each iteration, see related works from the literature for more details. The computational results are recorded over time at a given distance from the origin.

The same problem is solved numerically based on the mKdV model \eqref{mKdV} using 1024 points per period and a Fourier pseudo-spectral method, which is adapted from Trefethen's code for the KdV equation \citep[Prog. 27]{trefethen00}. To avoid numerical instability near discontinuities, we add a numerical viscosity term of the form $10^{-8} v_{tt}$ in the right-hand side of Eq.~\eqref{mKdV} before performing the pseudo-spectral simulations. The spatial step is set to $7.81 \times 10^{-5}$ m. The algorithm is iterated up to a given distance from the origin where the computational results are recorded.

\begin{figure*}
	\centering
	\begin{minipage}{0.49\linewidth}
		\centering
		
		$y = y^\star$
		
		\vspace{0.5em}
		
		\includegraphics{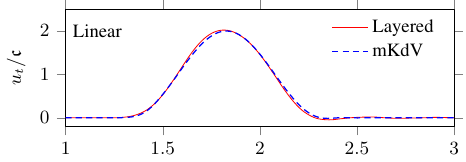}
		
		\includegraphics{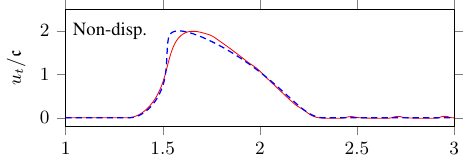}
		
		\includegraphics{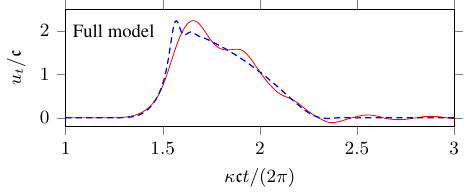}
	\end{minipage}\hfill
	\begin{minipage}{0.49\linewidth}
		\centering
		
		$y = 2y^\star$
		
		\vspace{0.5em}
		
		\includegraphics{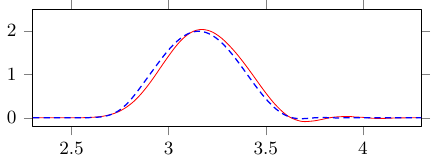}
		
		\includegraphics{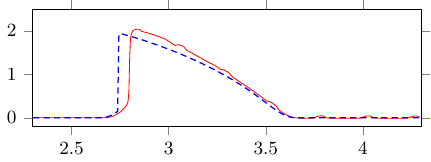}
		
		\includegraphics{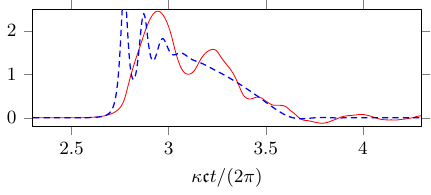}
	\end{minipage}

	\caption{Impact problem \eqref{Impact}. The numerical results obtained for the layered medium (solid lines) are compared to those obtained for the homogenised mKdV model \eqref{mKdV} (dashed lines), where the numerical solution has been recorded at two positions (left and right columns). The first row shows the linear dispersive solution, the second row shows the nonlinear non-dispersive solution, and the third row shows the nonlinear dispersive solution. \label{fig:Valid}}
\end{figure*}

The numerical results are shown in Figure~\ref{fig:Valid} for several propagation distances. Three cases are considered:
\begin{enumerate}
\item[(i)] the linear dispersive solution for which the parameters $\beta^{(\alpha)}$ are chosen equal to zero (i.e., $\zeta = 0$).
\item[(ii)] the nonlinear non-dispersive solution for which the acoustic impedance $\rho^{(2)}\mathfrak{c}^{(2)}$ is made equal to $\rho^{(1)}\mathfrak{c}^{(1)}$ by setting $\rho^{(2)} = 4650$~kg/m\textsuperscript{3} (this way, $\eta=0$). As a result, the numerical values of the effective wave speed $\mathfrak c$ and of the impact amplitude $\mathscr{V} = 2\mathfrak{c}$ are modified.\footnote{The mass density $\rho^{(2)} = \rho^{(1)}\mathcal{G}^{(1)}/\mathcal{G}^{(2)}$ is too large to be representative of a real elastomer. Hence, this fictional test case shall be understood as a thought experiment.\label{foot:Disc}}
\item[(iii)] the nonlinear dispersive solution that corresponds to all the parameter values provided in Table~\ref{tab:Param}.
\end{enumerate}
For each case, the heterogeneous model approximated by means of the finite volume method is compared to the homogenised mKdV model, which is discretised using the pseudo-spectral method.

In the first case (i), wave dispersion due to reflections at the layer interfaces manifests itself in the form of oscillations as the wave propagates. We observe that the mKdV simulations lead to faster oscillations than the heterogeneous simulation, which is consistent with the fact that the one-way theory allows higher-frequency waves to propagate (see Fig.~\ref{fig:Disp}). Nevertheless, the low-frequency bell-shaped waveform that propagates in the heterogeneous medium is well-captured by the mKdV model.

In the second case (ii), the elastic shock formation distance can be deduced from Eqs.~\eqref{mKdV}-\eqref{Impact}, which reduces to the initial-value problem for a nonlinear conservation law. Here, we find
\begin{equation}
	y^\star = \frac{16\sqrt 3}{9 \zeta\kappa} \frac{\mathfrak{c}^2}{\mathscr{V}^2} \approx 21~\text{cm} ,
	\label{ShockDist}
\end{equation}
see the methodology in the monograph by \citet{lax73}. In effect, we observe that a discontinuity develops in the mKdV solution at this particular propagation distance, but no such discontinuity is observed in the heterogeneous simulation at $y=y^\star$. Instead, shock formation takes place at a slightly larger propagation distance, as can be seen in simulation results recorded at $y=2 y^\star$ (this feature was already reported for homogeneous materials \citep{berjamin23}). The distortion of the wave at large amplitudes differs a little from one solution to the other, but the waveforms are qualitatively similar.

Finally, in the third case (iii), the effects of wave dispersion and of nonlinearity are combined. We observe that the mKdV solution and the heterogeneous solution exhibit higher-frequency oscillations and wave distorsion. More specifically, it seems that the wave breaks into a sequence of solitary waves. However, the resulting waveforms are quite dissimilar for the selected loading signal \eqref{Impact}. Looking back at our analysis of solitary wave solutions, we note that the maximum amplitude of such waves must satisfy
\begin{equation}
	\max u_t/\mathfrak{c} = \max \delta \mathfrak{s}/\mathfrak{c} \approx 2.7 ,
\end{equation}
in theory, see Eq.~\eqref{SolitonDeltMax}. While the maximum particle velocity in the heterogeneous solution remains below this critical value ($\max u_t/\mathfrak{c} \approx 2.5$), it exceeds it in the mKdV solution ($\max u_t/\mathfrak{c} \approx 3.0$). Thus, it appears that the homogenised model \eqref{WaveEffWautier} is more coherent with the computational results obtained for the heterogeneous medium than the mKdV model \eqref{mKdV}. Given the differences between the impact simulation and our single-soliton analysis, this comparison may give only a qualitative indication.

Based on all these observations, we conclude that the mKdV model should be used with a lot of care, especially for high-frequency waves of large amplitude. The reduced wave equation \eqref{mKdV} yields qualitative agreement with respect to the heterogeneous medium, but quantitative agreement is not achieved due to a poor representation of the linear dispersion features (Fig.~\ref{fig:Disp}), in general. Nevertheless, the mKdV equation is rather accurate for the modelling of unidirectional wave propagation in heterogeneous media with low dispersion at moderate amplitudes, as shown in Fig.~\ref{fig:Soliton} where solitary wave solutions are compared.

\begin{figure*}
	\centering
	\begin{minipage}{0.49\linewidth}
		\centering
		
		$y = y^\star$
		
		\vspace{0.5em}
		
		\includegraphics{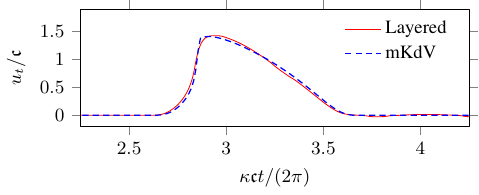}
	\end{minipage}\hfill
	\begin{minipage}{0.49\linewidth}
		\centering
		
		$y = 2y^\star$
		
		\vspace{0.5em}
		
		\includegraphics{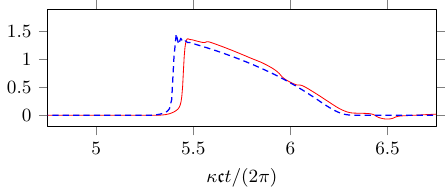}
	\end{minipage}

	\caption{Impact problem \eqref{Impact} for low dispersion and moderate amplitudes. The numerical results obtained for the layered medium (solid lines) are compared to those obtained for the homogenised mKdV model \eqref{mKdV} (dashed lines), where the numerical solution has been recorded at two positions (left and right columns). \label{fig:ValidKdV}}
\end{figure*}

To illustrate the relative accuracy of the mKdV model in some situations, we solve the impact problem for the layered medium and the mKdV model. Here, the parameters are chosen in such a way that the effects of wave dispersion and of large wave amplitudes are reduced. To keep the shock formation distance \eqref{ShockDist} unchanged, we set $\mathscr{V} = \mathfrak{c} \sqrt{2}$ and $2\pi/\kappa = 8\ell$ in the expression of the impact velocity \eqref{Impact}, which is still satisfactory from the point of view of scale separation. Furthermore, since the shear moduli $\mathcal{G}^{(\alpha)}$ and parameters of nonlinearity $\beta^{(\alpha)}$ are kept unchanged (cf. Table~\ref{tab:Param}), the parameter $\zeta$ is not modified either. Wave dispersion effects are reduced by setting $\rho^{(2)} = 4 \rho^{(1)} = 3720$~kg/m\textsuperscript{3} (this way, $\eta=9.26\times 10^{-5}$), which determines the numerical value of the effective wave speed $\mathfrak c${\,---\,}this value of the mass density $\rho^{(2)}$ is not meant to be representative of a real elastomer.\textsuperscript{\ref{foot:Disc}} As shown in Figure~\ref{fig:ValidKdV}, we observe that both solutions are very close over short propagation distances and times.

In the limit of small amplitude waves, numerical simulations based on a linear wave equation with mixed partial derivatives \eqref{WaveEffWautier} can be found in \citet{fish02}, as well as analytical results obtained for the linearised wave equation \eqref{WaveEff}. These results suggest that the full homogenised model is in good quantitative agreement with the heterogeneous model. The study of the boundary-value problem \eqref{Impact} for our effective theory \eqref{WaveEffWautier} is an interesting perspective, although it might be quite challenging due to the presence of the high-order spatial and temporal derivatives \citep{yong03}. Furthermore, higher-order homogenised boundary conditions need to be introduced for improved accuracy \citep{cornaggia20}. Thus, the study of initial-value problems for this high-order wave equation would be of interest, for instance with a nonzero initial strain and zero initial velocity. In this context, the development of an efficient and stable numerical method requires some attention. To this aim, the interested reader is also referred to the numerical results obtained by \citet{engelbrecht11}, and to relevant literature references therein.

\section{Magneto-active response and influence of the microstructure}\label{sec:Tune}

\subsection{Magneto-elastic tunability}

Let us return to the relationship \eqref{StretchMag} between the stretch $\lambda$ in the laminate and the magnetic field variables, which are the magnetic induction $b$ and the effective remnant induction, $\check{b}^\text{r}$, see its definition in Eq.~\eqref{StretchCoeff}. For instance, let us consider Gent materials with identical parameters $\beta^{(\alpha)} = \beta$ in both phases, cf. Table~\ref{tab:Param} for such a configuration. In this case, Eq.~\eqref{StretchMag} becomes
\begin{equation}
	\frac{\lambda^{2}-\lambda^{-1}}{1-\beta(\lambda^2 + 2\lambda^{-1} - 3)} = \left(1 - \frac{\mu_0}{\breve \mu}\right) b_\text{n}^2 + b_\text{n} \check{b}^\text{r}_\text{n} ,
	\label{StretchTune}
\end{equation}
where we have introduced
\begin{equation}
	b_\text{n} = \frac{b}{\sqrt{\mu_0\overline{\mathcal G}}}, \quad \check{b}^\text{r}_\text{n} = \frac{2 \check{b}^\text{r} \mu_0 / \breve \mu}{\sqrt{\mu_0\overline{\mathcal G}}} .
	\label{MagTune}
\end{equation}
Here, the stretching ratio can be expressed as a function of the magnetic induction. More precisely, $\lambda$ is a real root of a cubic equation, whose analytical expression can be written in closed form. Depending on the value of the parameters in the above equation, the stretch can either be a monotonous (increasing or decreasing) function of the normalised magnetic induction $b_\text{n}$, or a non-monotonous one \citep{zhang22}.

According to our homogenisation result \eqref{WaveEff}, a modification of the stretch $\lambda$ can affect the effective parameter $\eta$ governing wave dispersion, see the expressions in Tables~\ref{tab:Effective}-\ref{tab:Elast}. In fact, we have
\begin{equation}
	\eta = \frac{(\nu^{(1)} \nu^{(2)})^2}{12\, \langle \rho\rangle^2} \frac{( \rho^{(1)}G^{(1)} - \rho^{(2)}G^{(2)} )^2}{(\nu^{(1)}G^{(2)} + \nu^{(2)}G^{(1)})^2} ,
	\label{EtaTune}
\end{equation}
where the coefficients $G^{(\alpha)}$ of Table~\ref{tab:Elast} are evaluated at $I_1 = \lambda^2 + 2/\lambda$ (the other coefficients do not depend on $\lambda$). This general observation might have direct consequences on the evolution of the band gap frequencies \eqref{BandGap} and on properties of the solitary wave solutions (Section~\ref{sec:NLWaves}), which we discuss hereinafter.

For neo-Hookean laminates where the generalised shear moduli $G^{(\alpha)}$ are constant, we observe that $\eta$ and $\ell/\mathfrak{c}$ are stretch-independent. Therefore, the band gap frequencies defined in Eq.~\eqref{BandGap} are not tunable in this special case. This result is consistent with the exact dispersion relationship \eqref{DispExact}, which is stretch-independent in the neo-Hookean limit \citep{zhang22}. It is a consequence of the proportionality of the wave speeds $\mathfrak{c}$ and $\mathfrak{c}^{(\alpha)}$ with respect to $\lambda$, along with that of the characteristic lengths $\ell = \lambda L$ and $\ell^{(\alpha)} = \lambda L^{(\alpha)}$. 

For Gent laminates with identical parameters $\beta^{(\alpha)} = \beta$ in both phases, the generalised shear moduli $G^{(\alpha)}$ are no longer stretch-independent, but they have the same evolution with respect to $\lambda$. A consequence of this evolution is that $\eta$ does not depend on $\lambda$, and that the wave speed satisfies
\begin{equation}
	\mathfrak{c}^{(\alpha)} = \lambda \sqrt{ \frac{\mathcal{G}^{(\alpha)} / \rho^{(\alpha)}}{1 - \beta (\lambda^2 + 2\lambda^{-1}-3)} } .
	\label{SpeedTune}
\end{equation}
Thus, $\mathfrak{c}^{(\alpha)}$ is not proportional to $\lambda$, unless the phases have a neo-Hookean elastic response ($\beta=0$).

\begin{figure*}
	\centering
	\begin{minipage}{0.49\textwidth}
		\centering
		(a)
		
		\includegraphics{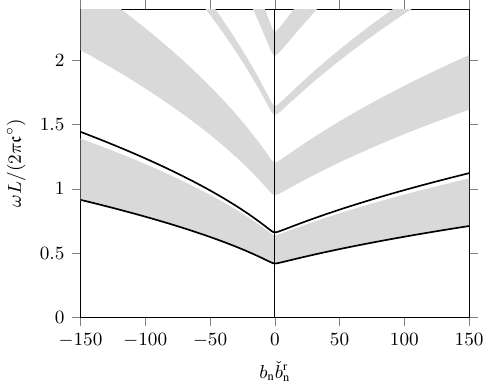}
	\end{minipage}\hfill
	\begin{minipage}{0.49\textwidth}
		\centering
		(b)
		
		\includegraphics{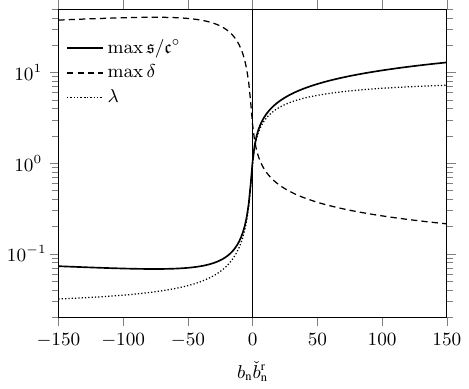}
	\end{minipage}
	\caption{(a) Evolution of the frequency band gaps from Fig.~\ref{fig:Disp} with respect to a normalised magnetic induction \eqref{MagTune}. The physical parameters of the Gent laminate are taken from Table~\ref{tab:Param}, and we have set $\breve\mu = \mu_0$. Shaded areas mark the band gaps deduced from the exact dispersion relationship \eqref{DispExact}, whereas black lines are deduced from the homogenised theory \eqref{BandGap}. (b) Evolution of the maximum velocity \eqref{MaxS} (solid line) and of the maximum strain amplitude \eqref{MaxStrain} (dashed line) of solitary wave solutions in the laminate. Evolution of the stretch according to the homogenised material theory (dotted line). \label{fig:TuneSoliton}}
\end{figure*}

We illustrate the tunability of the band gaps for such a Gent laminate in Figure~\ref{fig:TuneSoliton}a, where the evolution of the band gap frequencies in terms of the magnetic induction is shown. Here, we have set $\breve \mu = \mu_0$ which greatly simplifies Eq.~\eqref{StretchTune}, while the other parameters are taken from Table~\ref{tab:Param}. The angular frequency is rescaled with respect to the characteristic time $\ell/\mathfrak{c}$ measured in the undeformed material, where it is equal to the ratio $L/\mathfrak{c}^\circ$ with $\mathfrak{c}^\circ = \mathfrak{c}|_{\lambda=1}$. Thus,
\begin{equation}
	\frac{\omega L}{\mathfrak{c}^\circ} = \frac{\omega \ell / \mathfrak{c}}{\sqrt{1 - \beta (\lambda^2 + 2\lambda^{-1}-3)}} .
\end{equation}
Again, we note that the asymptotic homogenisation procedure yields a satisfactory estimation of the first band gap, whose cut-off frequencies depend on the magnetic induction. As displayed in Fig.~\ref{fig:TuneSoliton}b, the stretches deduced from \eqref{StretchTune} increase from 0.032 to 7.2 for $b_\text{n} \check{b}^\text{r}_\text{n}$ increasing from $-150$ to $150$. In other words, the material is extended along the lamination direction when $b_\text{n} \check{b}^\text{r}_\text{n}$ is increased.

Let us now investigate the tunability of the solitary wave solutions presented in Section~\ref{sec:NLWaves}, which do not exist in purely neo-Hookean laminates. For the homogenised wave equation \eqref{WaveEffWautier} with the parameters \eqref{WautierParam}, we have shown that the speed $\mathfrak s$ of a solitary wave is restricted by the upper bound $\mathfrak{s}^2/\mathfrak{c}^2 < \sqrt{1 + \eta/\eta_t}$. In addition, this maximum wave speed corresponds to a given maximum strain amplitude $\delta$, see Fig.~\ref{fig:Soliton} for an illustration.

For Gent laminates with identical $\beta^{(\alpha)} = \beta$, the maximum velocity of solitary waves is determined from
\begin{equation}
	\max \mathfrak{s}/\mathfrak{c}^\circ = \frac{\lambda \sqrt[4]{1 + \eta/\eta_t}}{\sqrt{1 - \beta (\lambda^2 + 2\lambda^{-1}-3)}} ,
	\label{MaxS}
\end{equation}
based on earlier definitions. Furthermore, the critical strain amplitude obtained from Eqs.~\eqref{SolitonWav}-\eqref{Soliton} for this special value of $\mathfrak{s}^2/\mathfrak{c}^2$ satisfies
\begin{equation}
	\frac{\max \delta}{(\sqrt{1 + \eta/\eta_t} - 1)^{1/2}} = \frac{ \sqrt{1-\beta (\lambda^2 + 2\lambda^{-1}-3)}}{\lambda\, (\frac12 \breve{\mathcal{G}}^3 \overline{\mathcal{G}^{-3}\beta} )^{1/2} } ,
	\label{MaxStrain}
\end{equation}
see Tables~\ref{tab:Effective}-\ref{tab:Elast} for the link between $\zeta$ and $\beta$, and Eqs.~\eqref{GalichCoeff}-\eqref{AvNote} for relevant notations.
In the present case, it follows that both the upper bounds for $\mathfrak{s}$ and for $\delta$ depend on the stretching ratio, or on the magnetic variables $b_\text{n} \check{b}^\text{r}_\text{n}$ given Eq.~\eqref{StretchTune}. In addition, we note in passing that these critical values of $\mathfrak{s}$ and $\delta$ follow the inverse trend when $\lambda$ is varied.

We represent the evolution of the critical wave speed \eqref{MaxS} and amplitude \eqref{MaxStrain} of solitary waves in Figure~\ref{fig:TuneSoliton}b, where the same parameter values were used as in Fig.~\ref{fig:TuneSoliton}a. According to our homogenisation result, we observe that the maximum admissible wave speed and amplitude for solitary wave solutions follow a non-monotonous evolution. In fact, we note that $\max \mathfrak{s}/\mathfrak{c}^\circ$ is an increasing function of the stretching ratio for $\lambda > \frac{3\beta}{1+3\beta} \approx 0.038$, but a decreasing function elsewhere. Furthermore, while the effect of magneto-deformations on the band gap frequencies is nearly symmetric with respect to the magnetic variables (Fig.~\ref{fig:TuneSoliton}a), its effect on the stretching ratio and on solitary wave solutions is not symmetric at all (Fig.~\ref{fig:TuneSoliton}b).

\subsection{Influence of the volume fractions}

Let us investigate the influence of the microstructure on wave propagation properties. Contrary to Fig.~\ref{fig:TuneSoliton} where the material was only deformed, we now vary parameters of the microstructure to reveal the main design features of our magneto-active laminate. Here, we study the influence of the volume fraction $\nu^{(2)} = 1 - \nu^{(1)}$ on wave propagation properties, while all the other parameters remain unchanged. Setting $\breve{\mu} = \mu_0$ and $b_\text{n} \check{b}^\text{r}_\text{n} = 0$ in Eq.~\eqref{StretchTune}, we find $\lambda = 1$. Hence, the laminate is in the undeformed configuration. Given the parameters in Table~\ref{tab:Param}, we note that the shear modulus $\mathcal{G}^{(\alpha)}$ in the softer phase ($\alpha=2$) is five times smaller than that in the stiffer phase ($\alpha=1$).

\begin{figure*}
	\centering
	\begin{minipage}{0.49\textwidth}
		\centering
		(a)
		
		\includegraphics{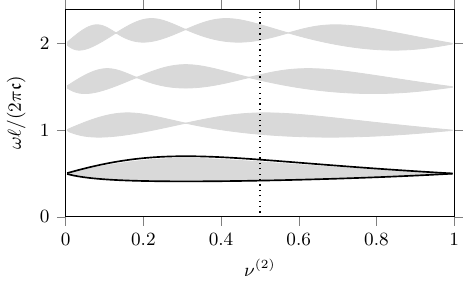} \\
		
		\vspace{0.2em}
		
		\includegraphics{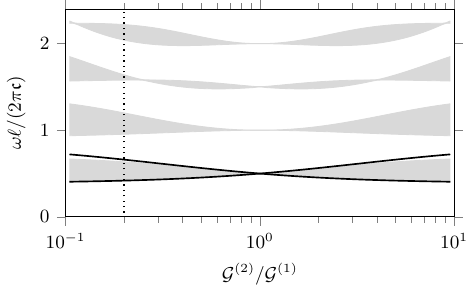}
	\end{minipage}\hfill
	\begin{minipage}{0.49\textwidth}
		\centering
		(b)
		
		\includegraphics{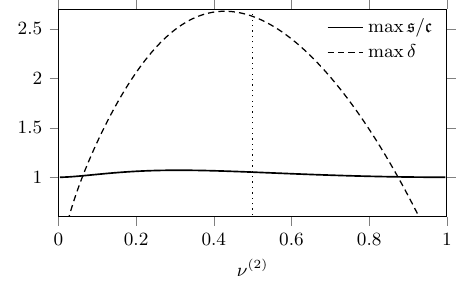} \\
		
		\vspace{0.2em}
		
		\includegraphics{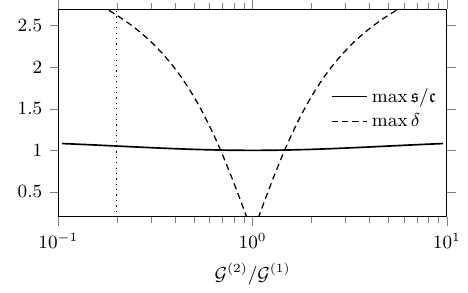}
	\end{minipage}
	\caption{Influence of the microstructure. (a) Evolution of the band gap frequencies with respect to the volume fraction $\nu^{(2)}$ (top), and with respect to the shear modulus contrast, $\mathcal{G}^{(2)}/\mathcal{G}^{(1)}$ (bottom). Shaded areas mark the band gaps deduced from the exact dispersion relationship \eqref{DispExact}, whereas black lines are deduced from the homogenised theory \eqref{BandGap}. (b) Maximum velocity \eqref{MaxS} and maximum strain amplitude \eqref{MaxStrain} of solitary waves in terms of the same parameters as in (a). Vertical dotted lines mark the parameter values in Table~\ref{tab:Param}. \label{fig:MicroSoliton}}
\end{figure*}

In Figure~\ref{fig:MicroSoliton}, we represent the evolution of the band gap frequencies with respect to the volume fraction $\nu^{(2)}$, as well as the evolution of the maximum velocity \eqref{MaxS} and of the maximum amplitude \eqref{MaxStrain} of solitary wave solutions. The general structure of the pass and stop bands in Fig.~\ref{fig:MicroSoliton}a is consistent with results obtained by \citet{andrianov08}, see Figure~S3 therein. According to these results, we note that some volume fractions lead to wider band gaps, and that our homogenised theory \eqref{BandGap} provides an accurate approximation of the exact dispersion relationship \eqref{DispExact} in the low frequency range. For instance, according to the homogenisation result \eqref{BandGap}, the first band gap is widest for the largest possible value of $\eta$, which is reached for the volume fraction
\begin{equation}
	\nu^{(2)} = \frac{\mathfrak{c}^{(2)}}{\mathfrak{c}^{(1)} + \mathfrak{c}^{(2)}} \approx 0.31 .
	\label{NuStar}
\end{equation}
However, as shown in the figure, the second and fourth band gaps are extremely narrow for this special value of $\nu^{(2)}$. Hence, for a given frequency, the filtering performance of such a laminate is very sensitive to the volume fractions.

Regarding the solitary wave solutions \eqref{Soliton}, we note that their propagation is facilitated in the range of volume fractions $0.2 < \nu^{(2)} < 0.5$. As shown in Fig.~\ref{fig:MicroSoliton}b, wave speeds and wave amplitudes can vary within a relatively wide range for these volume fractions. The maximum wave velocity \eqref{MaxS} reaches its largest value for the volume fraction in Eq.~\eqref{NuStar}, whereas the maximum wave amplitude \eqref{MaxStrain} reaches its largest value elsewhere ($\nu^{(2)} \approx 0.42$). In other words, perturbations with a variety of characteristic lengths and amplitudes will experience a soliton-like propagation within this range of volume fractions, for which the stiffer material is in excess of the softer one.

\subsection{Influence of the shear modulus contrast}

Let us now investigate the influence of the shear modulus contrast $\mathcal{G}^{(2)}/\mathcal{G}^{(1)}$ on wave propagation properties. Here, the laminate is undeformed, and all the other parameters are kept unchanged (see Table~\ref{tab:Param} for numerical values). As shown in Fig.~\ref{fig:MicroSoliton}a, the width of the frequency band gaps tends to increase with increasing or decreasing values of the shear modulus contrast away from $\mathcal{G}^{(2)}/\mathcal{G}^{(1)} = 1$. In Fig.~\ref{fig:MicroSoliton}b, the same observation can be made about the maximum velocity \eqref{MaxS} of solitary waves and about their maximum amplitude \eqref{MaxStrain}, which both increase for increasing or decreasing values of the shear modulus contrast away from $\mathcal{G}^{(2)}/\mathcal{G}^{(1)} = 1$. We note in passing that these curves are symmetric with respect to $\mathcal{G}^{(2)}/\mathcal{G}^{(1)} = 1$, given that all the other parameter values are the same for both phases.

\section{Conclusion}\label{sec:Conclu}

In Section~\ref{sec:Constitutive}, we have revisited the classical constitutive theory for hard-magnetic solids by \citet{zhao19}, by following \citet{dorfmann24}. This way, the usual symmetry requirements on the total Cauchy stress are satisfied. Then, this theory was used to describe nonlinear shearing motions in hard-magnetic soft laminates subjected to a permanent magnetic field along the lamination direction (Section~\ref{sec:Laminate}). Here, use was made of the asymptotic homogenisation procedure by \citet{andrianov13} to derive an effective wave equation \eqref{WaveEff}, as described in Section~\ref{sec:Homogen}. By combining this result with those from quasi-static homogenisation procedures \citep{debotton05}, it is possible to derive an effective strain energy function \eqref{Weff} for Yeoh laminates.

In Section~\ref{sec:NLWaves}, the modelling of linear wave dispersion is improved by following the steps in \citet{wautier15}. Then, the approximate mKdV equation \eqref{mKdV} is derived, and solitary wave solutions are studied. By performing direct numerical simulations based on finite volumes and pseudo-spectral methods \citep{yong03}, we show that the mKdV model is not quantitatively accurate, in general. More specifically, it fails to accurately represent wave dispersion, contrary to the full homogenised theory \eqref{WaveEffWautier} which nearly covers frequencies up to the second phononic band gap. The tunability and the optimisation of the laminate's properties is demonstrated in Section~\ref{sec:Tune}. In particular, we show that the band gaps can be adjusted in Gent laminates by varying the external magnetic field, the volume fractions, or the shear modulus contrast. A similar observation is made for the maximum admissible wave speed and amplitude of solitary wave solutions.

The main limitations of this study are direct consequences of its assumptions. Thus, let us recall that a hard-magnetic elastic theory for incompressible magneto-active solids was used, where the elastic response is assumed of generalised neo-Hookean type, see Eqs.~\eqref{StressIncomp}-\eqref{ElastEnergy}. Here, the influence of long-range self-magnetisation effects has been discarded. The layered medium has very large dimensions, and perfect bonding is assumed between the layers. Then, a permanent magnetic field in the lamination direction is applied \eqref{StretchMag}, and the static magneto-deformation is assumed homogeneous in each phase, which is only reasonable at a significant distance from the lateral boundaries. Then, plane shear waves propagating along the lamination direction are considered \eqref{FShear}, and fluctuations in the remnant magnetisation are neglected. Beyond these modelling assumptions, it is evident that the validity of our homogenised wave equation \eqref{WaveEffWautier} is restricted to the low frequency range and to waves of moderate amplitude, due to the assumptions of the homogenisation procedure and the postulated linearity of the dispersion terms. These limitations contrast with the validity of the exact dispersion equation \eqref{DispExact} (valid at any frequency and for small amplitudes), as well as with the validity of the finite-volume algorithm used in Section~\ref{subsec:Impact} (valid for any frequency and amplitude, in principle). It is important to keep these limitations in mind with regard to our analytical results on nonlinear wave propagation.

Several perspectives of the present study could become the scope of future investigations. In particular, the present hard-magnetic material model could be revised to incorporate long-range magnetic interactions \citep{steigmann04,rahmati23}, soft-magnetic behaviour \citep{mukherjee22,danas24} or hysteretic behaviour \citep{morro24,gebhart25}. This theory could also be exploited in a variety of applications, including the modelling of actuators and soft active structures such as rods and plates \citep{wang20,yan23}. In addition, various properties of the effective strain energy function \eqref{Weff} for Yeoh laminates could be analysed, such as the development of macro- and micro-structural instabilities \citep{rudykh13,yao24}.

In complement, it would be interesting to revisit the propagation of linear and nonlinear shear waves at oblique incidence angles by means of homogenisation theory, as well as the case of laminates with imperfect bonding between the layers \citep{alam25,chaki24}, or with finite spatial dimensions \cite{cornaggia20,marigo17}. Furthermore, the derivation of a more general three-dimensional effective theory could be attempted, which would encompass hard-magnetic materials as well nonlinear laminates with more general elastic responses (see the steps in \cite{spinelli15}). A key challenge in homogenisation theory is the accurate modelling of wave dispersion, which could be approached by an improved homogenised theory, nonlocal models \citep{willis09}, or by suitable generalised continuum theories \citep{madeo15, forest20}. In addition, the numerical resolution of nonlinear and dispersive wave equations of the form \eqref{WaveEffWautier} is a subject of interest \citep{engelbrecht11,cornaggia23}. Together with the finite-volume code used in Section~\ref{subsec:Impact}, such a computational framework would enable further validation of our homogenisation results.

%
%
%
%
%
\section*{Acknowledgements}

This project has received funding from the European Research Council (ERC) through Grant No. 852281 -- MAGIC.

\appendix

\section{Two-scale asymptotic analysis (details)}\label{app:Homogen}

In a similar fashion to \citet{andrianov13}, we describe the solution for $\text{u}_1$ valid at moderate wave amplitudes{\,---\,}the unknown fields $\text{u}_2$, $\text{u}_3$ are obtained in the limit of infinitesimal wave amplitudes, cf. relevant literature \citep{andrianov08,cornaggia20}. The field $\text{u}_1 = \text{U}_1 + \delta^2 \text{V}_1$ is sought as a truncated power series of the wave amplitude parameter $\delta$, where $\text{U}_1$ corresponds to the linear solution and $\text{V}_1$ is a correction accounting for nonlinearity.
In the nonlinear wave equation \eqref{WaveEps}-\eqref{Power}, the coefficient of order $\epsilon^{-1}$ requires that
\begin{equation}
	\begin{aligned}
	&\partial_{\tilde{\text y}\tilde{\text y}} \text{U}_1 = 0, & & (\text{order } \delta^0)\\
	&\partial_{\tilde{\text y}\tilde{\text y}} \text{V}_1 = 0 , & & (\text{order } \delta^2) 
	\end{aligned}
\end{equation}
together with the continuity of $\text{U}_1$, $\text{V}_1$ at material interfaces, which is deduced from the coefficient of order $\epsilon^{1}$ in \eqref{StressEps}\textsubscript{1}. From the coefficient of order $\epsilon^{0}$ in \eqref{StressEps}\textsubscript{2}, we deduce the continuity of
\begin{equation}
	\begin{aligned}
	&\text{g}^{(\alpha)} (\partial_{\text y} \text{u}_0 + \partial_{\tilde{\text y}} \text{U}_1) , & & (\text{order } \delta^0)\\
	&\text{g}^{(\alpha)} \partial_{\tilde{\text y}} \text{V}_1 + \tfrac13 \text{h}^{(\alpha)} (\partial_{\text y} \text{u}_0 + \partial_{\tilde{\text y}} \text{U}_1)^3 , & & (\text{order } \delta^2) 
	\end{aligned}
\end{equation}
at material interfaces.

The solution for $\text{U}_1$ is described in other works, and it reads
\begin{equation}
	\text{U}_1 = \tau \text{P} (\tilde{\text y} + \varphi) (\partial_{\text y}\text{u}_0) , \quad \text{P} = \frac{\text{g}^{(2)}-\text{g}^{(1)}}{\nu^{(1)} \text{g}^{(2)} + \nu^{(2)} \text{g}^{(1)}},
\end{equation}
where
\begin{equation}
	(\tau,\varphi) = \begin{cases}
		(\nu^{(2)}, \tfrac12), & -\tfrac12 \leq \tilde{\text y} \leq -\tfrac12\nu^{(2)}, \\
		(-\nu^{(1)}, 0) , & |\tilde{\text y}| \leq \tfrac12\nu^{(2)}, \\
		(\nu^{(2)}, -\tfrac12) , & \tfrac12\nu^{(2)} \leq \tilde{\text y} \leq \tfrac12 . \\
	\end{cases}
\end{equation}
Similarly, the solution for $\text{V}_1$ takes the form
\begin{equation}
	\text{V}_1 = \tau \text{Q} (\tilde{\text y} + \varphi) (\partial_{\text y} \text{u}_0)^3 , \;
	\text{Q} = \frac{\text{h}^{(2)} (\text{g}^{(1)})^3 - \text{h}^{(1)} (\text{g}^{(2)})^3}{3 \, (\nu^{(1)} \text{g}^{(2)} + \nu^{(2)} \text{g}^{(1)})^4} ,
\end{equation}
with the same coefficients $(\tau,\varphi)$ as for $\text{U}_1$. Up to the exponents in the expression of $\text{V}_1$, this expression is very similar to the one obtained by \citet{andrianov13} for materials with a quadratic nonlinearity in the stress-strain relationship. Finally, the leading-order homogenised wave equation is obtained by integrating the following equation with respect to the fast spatial coordinate over the unit cell $|\tilde{\text y}| \leq \frac12$,
\begin{equation}
	\begin{aligned}
	& \left(\text{g}^{(\alpha)} + \delta^2 \text{h}^{(\alpha)} (\partial_\text{y} \text{u}_0 + \partial_{\tilde{\text y}} \text{U}_1)^2\right) (\partial^2_\text{y} \text{u}_0 + \partial_{\text{y}\tilde{\text y}} \text{U}_1) \\
	&\qquad + \delta^2 \text{g}^{(\alpha)} \partial_{\text{y}\tilde{\text y}} \text{V}_1 = \varrho^{(\alpha)} \partial_\text{tt} \text{u}_0 ,
	\end{aligned}
\end{equation}
with $\alpha=2$ for $|\tilde{\text y}| \leq \frac12\nu^{(2)}$, and with $\alpha=1$ otherwise. The above identity corresponds to the coefficient of order $\epsilon^0$ in \eqref{WaveEps}-\eqref{Power}, to which we have subtracted the coefficient of order $\epsilon^1$ in $\partial_{\tilde{\text y}}$\eqref{StressEps}\textsubscript{2}. Then, we have used the continuity and periodicity requirements for \eqref{StressEps}\textsubscript{2} within the unit cell, and we have kept only leading order terms of the parameter $\delta$. The final result after integration consists of the $\epsilon^0$-order terms in Eq.~\eqref{WaveHom}.

\section{Towards a homogenised material theory (details)}\label{app:LopezPamies}

Let us consider a bi-laminate of inert generalised neo-Hookean materials \eqref{ElastEnergy} described by the strain energy function $W^{(\alpha)}$ which is function of the first strain invariant $I_1^{(\alpha)} = \bm{F}^{(\alpha)} : \bm{F}^{(\alpha)}$ in each phase ($\alpha = 1,2$), where the deformation gradient tensor $\bm{F}^{(\alpha)}$ is assumed uniform. To ensure that the displacement is continuous across the material interfaces, we require $\bm{F}^{(2)} \bm{m} = \bm{F}^{(1)} \bm{m}$, for any unit vector $\bm m$ orthogonal to the initial lamination direction indicated by the unit vector $\bm n$. This property allows us to represent the deformation gradients as
\begin{equation}
	\bm{F}^{(2)} - \bm{F}^{(1)} = \bm{\theta} \otimes \bm{n} , \quad \bm{\theta} = (\bm{F}^{(2)} - \bm{F}^{(1)}) \bm{n} .
\end{equation}
Introducing the macroscopic deformation $\bm{F} = \nu^{(1)}\bm{F}^{(1)} + \nu^{(2)}\bm{F}^{(2)}$, it follows that \citep{debotton05}
\begin{equation}
	\bm{F}^{(\alpha)} = \bm{F} + \tau^{(\alpha)} \bm{\theta} \otimes \bm{n} , \quad
	\tau^{(\alpha)} = \begin{cases}
	-\nu^{(2)}, \;\text{if}\; \alpha = 1,\\
	\phantom{-}\nu^{(1)}, \;\text{if}\; \alpha = 2.
	\end{cases}
	\label{FdeBotton}
\end{equation}
To enforce incompressibility in each phase, we require
\begin{equation}
	\det {\bm F}^{(\alpha)} = (\det {\bm F}) (1 + \tau^{(\alpha)} \bm{F}^{-1}\bm{\theta} \cdot \bm{n}) = 1 ,
\end{equation}
from which we deduce that $\bm{F}^{-1}\bm{\theta} \cdot \bm{n} = 0$. The vector $\bm{\theta}$ is determined based on the traction continuity equation $\bm{P}^{(2)} \bm{n} = \bm{P}^{(1)} \bm{n}$, with the tractions\footnote{The general expression of $(\bm{F}^{(\alpha)})^{-\text{T}}$ can be found in the Appendix of \citet{lopezpamies09}.}
\begin{equation}
	\bm{P}^{(\alpha)}\bm{n} = G^{(\alpha)} (\bm{F}\bm{n} + \tau^{(\alpha)} \bm{\theta}) - p^{(\alpha)} \bm{F}^{-\text{T}} \bm{n} .
\end{equation}
Here we have introduced the generalised shear moduli $G^{(\alpha)} = 2 W^{(\alpha)}_1$, as well as the Lagrange multipliers $p^{(\alpha)}$ to enforce incompressibility in each phase.

The pressure jump 
\begin{equation}
	p^{(2)}-p^{(1)} = \frac{G^{(2)} - G^{(1)}}{|\bm{F}^{-\text{T}}\bm{n}|^2} ,
\end{equation}
is deduced from the traction continuity equation upon scalar multiplication by $\bm{F}^{-\text{T}}\bm{n}$. After substitution in the difference $(\bm{P}^{(2)}-\bm{P}^{(1)})\bm{n}$, the Lagrange multipliers $p^{(\alpha)}$ can be eliminated from the traction continuity equation. The remaining terms yield the relationship
\begin{equation}
	\bm{\theta} = P \left( \bm{F}\bm{n} - \frac{\bm{F}^{-\text{T}} \bm{n}}{|\bm{F}^{-\text{T}}\bm{n}|^2} \right) , \quad
	P = \breve{G} \frac{G^{(1)}-G^{(2)}}{G^{(1)}G^{(2)}} ,
	\label{Spinelli}
\end{equation}
where $\breve{G} = (\nu^{(1)}/G^{(1)} + \nu^{(2)}/G^{(2)})^{-1}$ is the harmonic average of the generalised shear moduli. 
This expression of $\bm \theta$ is then used to calculate $I_1^{(\alpha)}$ from \eqref{FdeBotton}, which must satisfy
\begin{equation}
	I_1^{(\alpha)} = I_1 + \tau^{(\alpha)} P \left(2  + \tau^{(\alpha)} P\right) K,
	\label{INonlin}
\end{equation}
where $K$ was introduced in Eq.~\eqref{Galich}.
Finally, the effective strain energy function $W^\text{eff}$ is obtained by evaluating the spatial average $\nu^{(1)} W^{(1)} + \nu^{(2)} W^{(2)}$. In the special case of neo-Hookean solids for which $W^{(\alpha)} = \frac12 \mathcal{G}^{(\alpha)} (I_1^{(\alpha)} - 3)$, we recover the expression \eqref{Galich}.

We consider now the Yeoh, Fung-Demiray and Gent models. For small values of the parameter $\beta^{(\alpha)}$, all these strain energy functions are equivalent to that from the two-term Yeoh theory, thus we will restrict the study to this particular model{\,---\,}this observation was also exploited by \citet{shmuel10} in a similar fashion for the study of fibre-reinforced composites, see also related works on the subject \citep{agoras09, berjamin25}. Injecting the expression of $G^{(\alpha)}$ from Table~\ref{tab:Elast} into Eqs.~\eqref{Spinelli}\textsubscript{2}-\eqref{INonlin}, a nonlinear algebraic system of equations is obtained in terms of the invariants $I_1^{(\alpha)}$, in general. Here, we solve these equations in an approximate fashion by linearisation of the equations and of their solution using Taylor series expansions valid for $\beta^{(\alpha)} \ll 1$.

At leading order in $\beta^{(1)}$ and $\beta^{(2)}$, we obtain a polynomial expression for the energy (using symbolic calculus software), which takes the form
\begin{equation}
	\begin{aligned}
	W^\text{eff} &= \tfrac12 \overline{\mathcal G} (I_1-3) - \tfrac12 (\overline{\mathcal G} - \breve{\mathcal G}) K \\
	&\quad + \Gamma_{20} (I_1-3)^2  + \Gamma_{11} (I_1-3) K + \Gamma_{02} K^2 ,
	\end{aligned}
	\label{WGen}
\end{equation}
consistently with the exact formula \eqref{Galich}. The coefficients $\Gamma_{ij}$ can be determined as follows:
\begin{itemize}
	\item[(i)] the quasi-static response of the laminate in tension-compression yields the average energy $W^\text{eff} = \overline{W}$ for $K=0$, so that $\Gamma_{20} = \frac14 \overline{\mathcal{G}^1\beta}$, where we have used the notation \eqref{AvNote} for the coefficients $\overline{\mathcal{G}^n\beta}$ with $n$ integer.
	\item[(ii)] the shearing motion \eqref{FShear} along the layer interfaces is governed by a wave equation that is consistent with our homogenisation result \eqref{WaveEff} in the elastic limit ($\eta = 0$), which yields the constraints
	\begin{equation}
		\begin{aligned}
			&\langle g \rangle = \lambda^2 \left(\breve{\mathcal G} + (4\Gamma_{20} + 2\Gamma_{11})(\mathcal{I}_1-3)\right) , \\
			&\langle g \rangle \zeta = 12\lambda^4 \left(\Gamma_{20} + \Gamma_{11} + \Gamma_{02}\right) .
		\end{aligned}
	\end{equation}
	In addition, we exploit the following linearised formulas deduced from Tables~\ref{tab:Effective}-\ref{tab:Elast} for small $\beta^{(\alpha)}$:
	\begin{equation}
		\begin{aligned}
			&\langle g \rangle \simeq \lambda^2 \left(\breve{\mathcal G} + \breve{\mathcal G}^2 \,\overline{ \mathcal{G}^{-1}\beta }\, (\mathcal{I}_1-3)\right) ,\\
			&\langle g \rangle \zeta \simeq 3\lambda^4 \breve{\mathcal G}^4 \,\overline{ \mathcal{G}^{-3}\beta } .
		\end{aligned}
	\end{equation}
\end{itemize}
This way, we have determined concise expressions for all the coefficients in Eq.~\eqref{WGen}. The final expression of the effective strain energy function is provided in Eq.~\eqref{Weff}.

\bibliographystyle{unsrtnat}
\bibliography{Biblio}{}

\end{document}